\documentstyle[aps,psfig,floats,amsfonts,eqsecnum]{revtex}

\begin{document}
\draft
\wideabs{
  
  \title{The influence of magnetic-field-induced spin-density-wave
    motion and finite temperature on the quantum Hall effect in
    quasi-one-dimensional conductors: A quantum field theory}

\author{Victor M.~Yakovenko\cite{Yakovenko} and Hsi-Sheng Goan\cite{Goan}}

\address{Department of Physics and Center for Superconductivity Research,
University of Maryland, College Park, MD 20742, USA}
\date{{\bf cond-mat/9804128}, April 12, 1998}
\maketitle
\begin{abstract}
  We derive the effective action for a moving magnetic-field-induced
  spin-density wave (FISDW) in quasi-one-dimensional conductors at
  zero and nonzero temperatures by taking the functional integral over
  the electron field.  The effective action consists of the (2+1)D
  Chern-Simons term and the (1+1)D chiral anomaly term, both written
  for a sum of the electromagnetic field and the chiral field
  associated with the FISDW phase.  The calculated frequency
  dependence of Hall conductivity interpolates between the quantum
  Hall effect at low frequencies and zero Hall effect at high
  frequencies, where the counterflow of FISDW cancels the Hall
  current.  The calculated temperature dependence of the Hall
  conductivity is interpreted within the two-fluid picture, by analogy
  with the BCS theory of superconductivity.

\end{abstract}

}

\section{INTRODUCTION} 
\label{sec:Intro}

Organic metals of the (TMTSF)$_2$X family, where TMTSF is
tetramethyltetraselenafulvalene and X represents an inorganic anion
such as ClO$_4$ or PF$_6$, are highly anisotropic,
quasi-one-dimensional (Q1D) crystals that consist of parallel
conducting chains (see reviews \cite{Yamaji90,Jerome94}).  The
electron wave functions overlap and the electric conductivity are the
highest in the direction of the chains (the {\bf a} direction) and are
much smaller in the {\bf b} direction perpendicular to the chains.  In
this paper, we neglect coupling between the chains in the third, {\bf
c} direction, which is weaker than in the {\bf b} direction, and model
(TMTSF)$_2$X as a system of uncoupled two-dimensional (2D) layers
parallel to the {\bf a}-{\bf b} plane, each of the layers having a
strong Q1D anisotropy.  We choose the coordinate axis $x$ along the
chains and the axis $y$ perpendicular to the chains within a layer.

A moderate magnetic field $H$ of the order of several Tesla, applied
perpendicular to the layers, induces the so-called
magnetic-field-induced spin-density wave (FISDW) in the system (see
review \cite{Chaikin96}).  In the FISDW state, the electron spin
density is periodically modulated along the chains with the wave
vector
\begin{equation}
  Q_x=2k_F-NG,
\label{eq:Qx}  
\end{equation}
where $k_F$ is the Fermi wave vector of the electrons, $N$ is an
integer that characterizes FISDW, and
\begin{equation}
  G=\frac{ebH}{\hbar c}
\label{eq:G(H)}
\end{equation}
is a characteristic wave vector of the magnetic field.  In Eq.\ 
(\ref{eq:G(H)}), $e$ is the electron charge, $\hbar=h/2\pi$ is the
Planck constant, $c$ is the speed of light, and $b$ is the distance
between the chains.  The longitudinal wave vector of FISDW
(\ref{eq:Qx}) is not equal to $2k_F$ (as it would in a purely
one-dimensional (1D) case), but deviates by an integer multiple of the
magnetic wave vector $G$.  When the magnetic field changes, the
integer $N$ stays constant within a certain range of the magnetic
field, then switches to another value, and so on.  Thus, the system
exhibits a cascade of the FISDW phase transitions when the magnetic
field changes.  The theory of FISDW was initiated by Gor'kov and
Lebed' \cite{Lebed84}, further developed in Refs.\ 
\cite{Lebed85,Montambaux84b,Montambaux85,Montambaux86,Montambaux88a,Yamaji85,Maki86a,Maki86b,Maki87b},
and reviewed in Refs.\ \cite{Montambaux91,Lederer96}.

Within each FISDW phase, the Hall conductivity per one layer,
$\sigma_{xy}$, has an integer quantized value at zero temperature:
\begin{equation}
  \sigma_{xy}=\frac{2Ne^2}{h},
\label{eq:2Ne2/h}
\end{equation}
where $N$ is the same integer that appears in Eq.\ (\ref{eq:Qx}) and
characterizes FISDW.  (The factor 2 in Eq.\ (\ref{eq:2Ne2/h}) comes
from the two orientations of the electron spin.)  A gap in the energy
spectrum of the electrons, which is a necessary condition for the
quantum Hall effect (QHE), is supplied by FISDW.  The theory of QHE in
the FISDW state of Q1D conductors was developed in Refs.\ 
\cite{Montambaux87,Yakovenko91a,Machida94} (see also reviews
\cite{Lederer96,Yakovenko96}).  The theory assumes that FISDW is
pinned and acts on electrons as a static periodic potential, so that
Eq.\ (\ref{eq:2Ne2/h}) represents QHE \cite{Thouless} in a 2D periodic
potential produced by FISDW and the chains.

On the other hand, under certain conditions, a density wave in a Q1D
conductor can move (see, for example, reviews \cite{Gruner88}).  It is
interesting to find out how this motion would affect QHE.  Since the
density-wave condensate can move only along the chains, at first
sight, this purely 1D motion cannot contribute to the Hall effect,
which is essentially a 2D effect.  Nevertheless, we show in this paper
that in the case of FISDW, unlike in the case of a regular charge- or
spin-density wave (CDW/SDW), a {\it nonstationary} motion of the FISDW
condensate does produce a nontrivial contribution to the Hall
conductivity.  In an ideal system, where FISDW is not pinned or
damped, this additional contribution due to the FISDW motion (the
so-called Fr\"{o}hlich conductivity \cite{Gruner88}) would exactly
cancel the bare QHE, so that the resultant Hall conductivity would be
zero.  In real systems, this effect should result in vanishing of the
ac Hall conductivity at high enough frequencies, where the dynamics of
FISDW is dominated by inertia, and pinning and damping can be
neglected.  Because we study an interplay between QHE and the
Fr\"{o}hlich conductivity, our theory has some common ideas with the
so-called topological superconductivity theory \cite{Wiegmann92},
which also seems to contain these ingredients.  Frequency dependence
of the Hall conductivity in a FISDW system was studied theoretically
in Ref.\ \cite{Maki89}.  However, because this theory fails to produce
QHE at zero frequency, it is unsatisfactory.  Some unsuccessful
attempts to derive an effective action for a moving FISDW and QHE were
made in Ref.\ \cite{Rozhavsky92}.

Another interesting question is how the Hall conductivity in the FISDW
state depends on temperature $T$. Our calculations show that thermal
excitations across the FISDW energy gap partially destroy QHE, and
$\sigma_{xy}(T)$ interpolates between the quantized value
(\ref{eq:2Ne2/h}) at zero temperature and zero value at the transition
temperature $T_c$, where FISDW disappears.  We find that
$\sigma_{xy}(T)$ has a temperature dependence similar to that of the
superfluid density in the BCS theory of superconductivity.  Thus, at a
finite temperature, one might think of a two-fluid picture of QHE,
where the Hall conductivity of the condensate is quantized, but the
condensate fraction of the total electron density decreases with
increasing temperature.  An attempt to calculate the Hall conductivity
in the FISDW state at a finite temperature was made in Ref.\ 
\cite{Maki89}, but it failed to produce QHE at zero temperature.

Some of our results were briefly reported in conference proceedings
\cite{Yakovenko93}.  They were also presented on a heuristic,
semiphenomenological level in our review \cite{Yakovenko96}.  In the
current paper, we present a systematic derivation of these results
within the quantum-field-theory formalism.  In Sec.\ 
\ref{sec:phenomenological}, we heuristically derive the effective
Lagrangian of a moving FISDW and the corresponding ac Hall
conductivity.  In Sec.\ \ref{sec:(1+1)DW}, as a warmup exercise, we
formally derive the effective action of a (1+1)D CDW/SDW in order to
demonstrate that our method reproduces well-known results in this
case.  In the quantum-field theory, this effective action is usually
associated with the so-called chiral anomaly
\cite{Krive85,Su86,Brazovskii93}.  Our method of derivation is close
to that of Ref.\ \cite{Ishikawa88}. In Sec.\ \ref{sec:2+1}, we
generalize the method of Sec.\ \ref{sec:(1+1)DW} to the case of (2+1)D
FISDW and derive the effective action for a moving FISDW.  We find
that, in addition to the (1+1)D chiral anomaly term, the effective
action contains the (2+1)D Chern-Simons term, written for a
combination of electromagnetic potentials and gradients of the FISDW
phase.  This modified Chern-Simons term describes both the QHE of a
static FISDW and the effect of FISDW motion.  The results are
consistent with the heuristic derivation of Sec.\ 
\ref{sec:phenomenological}.  In Sec.\ \ref{sec:alternative}, we
rederive the results of Sec.\ \ref{sec:2+1} using an alternative
method, which then is straightforwardly generalized to a finite
temperature in Sec.\ \ref{sec:temperature}.  The results for a finite
temperature are obtained heuristically in Sec.\ \ref{sec:Galileo} and
formally in Sec.\ \ref{sec:Matsubara}.  Experimental implications of
our theory are discussed in Sec.\ \ref{sec:experiment}.  Conclusions
are given in Sec.\ \ref{sec:conclusions}.

\section{SEMIPHENOMENOLOGICAL APPROACH TO QHE AND MOTION OF FISDW}
\label{sec:phenomenological}

\subsection{Fr\"{o}hlich current and Hall current}
\label{sec:current}

We consider a 2D system where electrons are confined to the chains
parallel to the $x$ axis, and the spacing between the chains along the
$y$ axis is equal to $b$.  A magnetic field $H$ is applied along the
$z$ axis perpendicular to the $(x,y)$ plane. The system is in the
FISDW state at zero temperature.  In order to calculate the Hall
effect, let us apply an electric field ${\cal E}_y$ perpendicular to
the chains. The electron Hamiltonian ${\Bbb H}$ can be written as
\begin{eqnarray}
  {\Bbb H}&=&-\frac{\hbar^2}{2m}\frac{\partial^2}{\partial x^2}
  +2\Delta\cos(Q_x x+\Theta) 
\nonumber \\
  && +2t_b\cos(k_y b-G x +\Omega_y t),
\label{eq:H}
\end{eqnarray}
where $k_y$ is the electron wave vector perpendicular to the chains.
In the r.h.s.\ of Eq.\ (\ref{eq:H}), the first term represents the
kinetic energy of the electron motion along the chains with the
effective mass $m$.  The second term describes the periodic potential
produced by FISDW.  The FISDW potential is characterized by the
longitudinal wave vector $Q_x$ (\ref{eq:Qx}), an amplitude $\Delta$,
and a phase $\Theta$.  The third term describes the electron tunneling
between the nearest neighboring chain with the amplitude $t_b$.  In
the gauge $A_y=Hx-c{\cal E}_yt$ and $\phi=A_x=A_z=0$, the magnetic and
the transverse electric fields appear in the third term of Hamiltonian
(\ref{eq:H}) via the Peierls-Onsager substitution $k_y\rightarrow
k_y-eA_y/c\hbar$, where $\Omega_y=eb{\cal E}_y/\hbar$, and $G$ is
given by Eq.\ (\ref{eq:G(H)}). Strictly speaking, a complete theory of
FISDW requires us to take into account the transverse component $Q_y$
of the FISDW wave vector and the electron tunneling between the
next-nearest neighboring chains with the amplitude $t_b'$
\cite{Montambaux84b,Montambaux85,Montambaux86,Montambaux88a}.
However, while $t_b'$ and $Q_y$ are very important for determining the
properties of FISDW, such as $N$, $\Delta$, and $T_c$, $t_b'$ and
$Q_y$ are not essential for the theory of QHE, so we set them at zero
in order to simplify presentation.  We do not pay attention to the
spin structure of the density-wave order parameter in Eq.\
(\ref{eq:H}), because it is immaterial for our study, which focuses on
the orbital effect of the magnetic field.  To simplify the
presentation, we study the case of CDW, but the results for SDW are
the same.

In the presence of the magnetic field $H$, the interchain hopping term
in Eq.\ (\ref{eq:H}) acts as a potential, periodic along the chains
with the wave vector $G$ proportional to $H$. In the presence of the
transverse electric field ${\cal E}_y$, this potential moves along the
chains with the velocity $\Omega_y/G=c{\cal E}_y/H$ proportional to
${\cal E}_y$. This velocity is nothing but the drift velocity in
crossed electric and magnetic fields. The FISDW potential may also
move along the chains, in which case its phase $\Theta$ depends on
time $t$, and the velocity of the motion is proportional to the time
derivative $\dot{\Theta}$. We are interested in a spatially
homogeneous motion of FISDW, so let us assume that $\Theta$ depends
only on time $t$ and not on the coordinates $x$ and $y$. We also
assume that both potentials move very slowly, adiabatically, which is
the case when the electric field is sufficiently weak.

Let us calculate the current along the chains produced by the motion
of the potentials. Since there is an energy gap at the Fermi level,
following the arguments of Laughlin \cite{Laughlin81} we can say that
an integer number of electrons $N_1$ is transferred from one end of a
chain to another when the FISDW potential shifts by its period
$l_1=2\pi/Q_x$. The same is true for the motion of the interchain
hopping potential with an integer $N_2$ and the period $l_2=2\pi/G$.
Suppose that the first potential shifts by an infinitesimal
displacement $dx_1$ and the second by $dx_2$. The total transferred
charge $dq$ would be the sum of the prorated amounts of $N_1$ and
$N_2$:
\begin{equation}
  dq=eN_1\frac{dx_1}{l_1}+eN_2\frac{dx_2}{l_2}.
\label{eq:dq}
\end{equation}
Now, suppose that both potentials are shifted by the same displacement
$dx=dx_1=dx_2$. This corresponds to a translation of the system as a
whole, so we can write that
\begin{equation}
  dq=e\rho\,dx,
\label{eq:rho}
\end{equation}
where $\rho=4k_F/2\pi$ is the concentration of electrons. Equating
(\ref{eq:dq}) and (\ref{eq:rho}) and substituting the expressions for
$\rho$, $l_1$, and $l_2$, we find the following Diophantine-type
equation \cite{Zak}:
\begin{equation}
  4k_F=N_1 (2 k_F-NG)+N_2 G.
\label{eq:Diophant}
\end{equation}
Since $k_F/G$ is, in general, an irrational number, the only
solution of Eq.\ (\ref{eq:Diophant}) for the integers $N_1$ and $N_2$ is
$N_1=2$ and $N_2=N_1 N=2N$.

Dividing Eq.\ (\ref{eq:dq}) by a time increment $dt$ and the
interchain distance $b$, we find the density of current along the
chains, $j_x$. Taking into account that according to Eq.\ (\ref{eq:H})
the displacements of the potentials are related to their phases:
$dx_1=-d\Theta/Q_x$ and $dx_2=\Omega_ydt/G$, we find the final
expression for $j_x$:
\begin{equation}
  j_x=-\frac{e}{\pi b}\dot{\Theta} + \frac{2Ne^2}{h}{\cal E}_y.
\label{eq:jx}
\end{equation}
The first term in Eq.\ (\ref{eq:jx}) represents the contribution of
the FISDW motion, the so-called Fr\"{o}hlich conductivity
\cite{Gruner88}. This term vanishes when the FISDW is pinned and does
not move ($\dot{\Theta}=0$).  The second term in Eq.\ (\ref{eq:jx})
describes QHE, in agreement with Eq.\ (\ref{eq:2Ne2/h}).

\subsection{Effective Lagrangian}
\label{sec:action}

To complete the solution of the problem, it is necessary to find how
$\dot{\Theta}$ depends on ${\cal E}_y$. For this purpose, we need the
equation of motion for $\Theta$, which can be derived once we know the
Lagrangian density of the system, $L$.  Two terms in $L$ can be
readily recovered taking into account that the current density $j_x$,
given by Eq.\ (\ref{eq:jx}), is the variational derivative of the
Lagrangian density with respect to the electromagnetic vector
potential $A_x$: $j_x=c\,\delta L/\delta A_x$.  Written in a
gauge-invariant form, the recovered part of the Lagrangian density is
equal to
\begin{equation} 
  L_1=\frac{Ne^2}{2\pi\hbar c}\varepsilon_{ijk}
  A^i\frac{\partial A^k}{\partial x_j}
  -\frac{e}{\pi b}\Theta {\cal E}_x,
\label{eq:L1}
\end{equation}
where the first term is the so-called Chern-Simons term responsible
for QHE \cite{Yakovenko91a}, and the second term describes the
interaction of the density-wave condensate with the electric field
along the chains ${\cal E}_x=-\partial A_x/c\partial
t-\partial\phi/\partial x$ \cite{Gruner88}.  In Eq.\ (\ref{eq:L1}), we
use the relativistic notation \cite{Landau-II} with the indices
$(i,j,k)$ taking the values $(0,1,2)$ and the implied summation over
repeated indices \cite{indices}.  The contravariant vectors have the
superscript indices: $A^i=(\phi,A_x,A_y)$ and $x^j=(ct,x,y)$.  The
covariant vectors have subscript indices: $x_j=(ct,-x,-y)$, and are
obtained from the contravariant vectors by applying the metric tensor
of the Minkowski space: $g_{ij}=g^{ij}={\rm diag}(1,-1,-1)$.
$\varepsilon_{ijk}$ is the antisymmetric tensor with
$\varepsilon_{012}=1$.  The potentials $A^i$ and the corresponding
fields ${\cal E}_x$, ${\cal E}_y$, and ${\cal H}_z$ represent an
infinitesimal external electromagnetic field.  These potentials do not
include the vector potential of the bare magnetic field $H$, which is
incorporated into the Hamiltonian of the system via the term $Gx$ in
Eq.\ (\ref{eq:H}) with $G$ given by Eq.\ (\ref{eq:G(H)}).

Lagrangian density (\ref{eq:L1}) should be supplemented with the
kinetic energy of the FISDW condensate, $K$.  The FISDW potential
itself has no inertia, because it is produced by the instantaneous
Coulomb interaction between electrons, so $K$ originates completely
from the kinetic energy of the electrons confined under the FISDW
energy gap.  Thus $K$ is proportional to the square of the average
electron velocity, which, in turn, is proportional to the electric
current along the chains:
\begin{equation}
  K=\frac{\pi\hbar b}{4v_Fe^2}\,j_x^2,                  
\label{eq:K}
\end{equation}
where $v_F=\hbar k_F/m$ is the Fermi velocity. Substituting Eq.\ 
(\ref{eq:jx}) into Eq.\ (\ref{eq:K}), expanding, and omitting an
unimportant term proportional to ${\cal E}_y^2$, we obtain the second
part of the Lagrangian density of the system:
\begin{equation}
  L_2=\frac{\hbar}{4\pi bv_F}\dot{\Theta}^2
  -\frac{eN}{2\pi v_F}\dot{\Theta}{\cal E}_y.                
\label{eq:L2}
\end{equation}
The first term in Eq.\ (\ref{eq:L2}) is the same as the kinetic energy
of a purely 1D density wave \cite{Gruner88} and is not specific to
FISDW.  The most important is the second term, which describes the
interaction of the FISDW motion and the electric field perpendicular
to the chains. This term is allowed by symmetry in the considered
system and has the structure of a mixed vector--scalar product:
\begin{equation}
  {\bf v} [\mbox{\boldmath${\cal E}$}\times{\bf H}].
\label{eq:vEH}
\end{equation}
Here, ${\bf v}$ is the velocity of the FISDW, which is proportional to
$\dot{\Theta}$ and is directed along the chains, that is, along the
$x$ axis.  The magnetic field {\bf H} is directed along the
$z$ axis, thus allowing the electric field {\boldmath${\cal E}$} to
enter only through the component ${\cal E}_y$.  Comparing Eq.\ 
(\ref{eq:vEH}) with the last term in Eq.\ (\ref{eq:L2}), one should
take into account that the magnetic field enters the last term
implicitly, through the integer $N$, which depends on $H$ and
changes sign when $H$ changes sign.

Varying the total Lagrangian $L=L_1+L_2$, given by Eqs.\ (\ref{eq:L1})
and (\ref{eq:L2}), with respect to $A_y$, we find the current density
across the chains:
\begin{equation}
  j_y=-\frac{2Ne^2}{h}{\cal E}_x-\frac{eN}{2\pi v_F}\ddot{\Theta}.
\label{eq:jy}
\end{equation}
In the r.h.s.\ of Eq.\ (\ref{eq:jy}), the first term describes the
quantum Hall current, whereas the second term, proportional to the
{\it acceleration} of the FISDW condensate, comes from the second term
in Eq.\ (\ref{eq:L2}) and reflects the contribution of the FISDW
motion along the chains to the electric current across the chains.

Setting the variational derivative of $L$ with respect to $\Theta$ to
zero, we find the equation of motion for $\Theta$:
\begin{equation}
  \ddot{\Theta}=-\frac{2ev_F}{\hbar}{\cal E}_x 
  + \frac{eNb}{\hbar}\dot{\cal E}_y.
\label{eq:EOM}
\end{equation}
In Eq.\ (\ref{eq:EOM}), the first two terms constitute the standard 1D
equation of motion of the density wave\cite{Gruner88}, whereas the
last term, proportional to the time derivative of ${\cal E}_y$, which
originated from the last term in Eq.\ (\ref{eq:L2}), describes the
influence of the electric field across the chains on the motion of
FISDW.

\subsection{Hall conductivity}
\label{sec:Hall}

In order to see the influence of the FISDW motion on the Hall effect,
let us consider the two cases, where the electric field is applied
either perpendicular or parallel to the chains.  In the first case,
${\cal E}_x=0$, so integrating Eq.\ (\ref{eq:EOM}) in time, we find
that $\dot{\Theta}=eNb{\cal E}_y/\hbar$.  Substituting this equation
into Eq.\ (\ref{eq:jx}), we see that the first term (the Fr\"{o}hlich
conductivity of FISDW) precisely cancels the second term (the quantum
Hall current), so {\it the resulting Hall current is equal to zero}.
This result could have been obtained without calculations by taking
into account that the time dependence $\Theta(t)$ is determined by the
principle of minimal action.  The relevant part of the action is
given, in this case, by Eq.\ (\ref{eq:K}), which attains the minimal
value at zero current: $j_x=0$.  We can say that if FISDW is free to
move it adjusts its velocity to compensate the external electric
field ${\cal E}_y$ and to keep zero Hall current.  In the second case,
where the electric field ${\cal E}_x$ is directed along the chains, it
accelerates the density wave according to the equation of motion
(\ref{eq:EOM}): $\ddot{\Theta}=-2ev_F{\cal E}_x/\hbar$.  Substituting
this equation into Eq.\ (\ref{eq:jy}), we find again that the Hall
current vanishes.

It is clear, however, that in stationary, dc measurements, the
acceleration of the FISDW, discussed in the previous paragraph, cannot
last forever.  Any friction or dissipation will inevitably stabilize
the motion of the density wave to a steady flow with zero
acceleration.  In this steady state, the second term in Eq.\ 
(\ref{eq:jy}) vanishes, and the current $j_y$ recovers its quantum
Hall value.  The same is true in the case where the electric field is
perpendicular to the chains.  In that case, dissipation eventually
stops the FISDW motion along the chains and restores $j_x$, given by
Eq.\ (\ref{eq:jx}), to the quantum Hall value. The conclusion is that
the contribution of the moving FISDW condensate to the Hall
conductivity is essentially nonstationary and cannot be observed in dc
measurements.

\begin{figure}
\centerline{\psfig {file=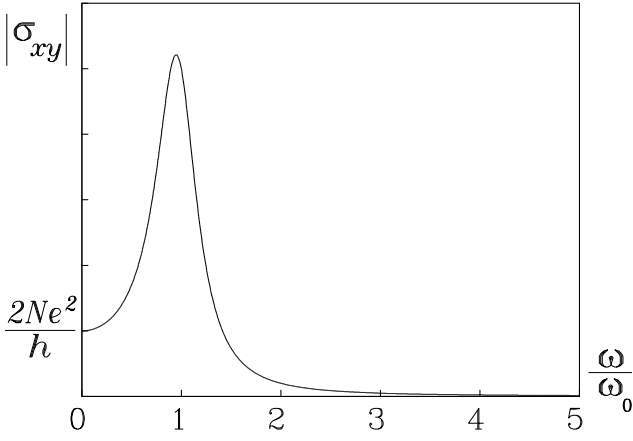,width=\linewidth,angle=-90}}
\caption{ Absolute value of the Hall conductivity in the FISDW state
as a function of the frequency $\omega$ normalized to the pinning
frequency $\omega_0$, as given by Eq.\ (\ref{eq:omega}) with
$\omega_0\tau=2$.}
\label{fig:Hall(omega)}
\end{figure}

On the other hand, the effect can be seen in ac experiments.  To be
realistic, let us add damping and pinning \cite{Gruner88} to the
equation of motion of FISDW (\ref{eq:EOM}):
\begin{equation}
  \ddot{\Theta}+\frac{1}{\tau}\dot{\Theta}+\omega_0^2\Theta
  =-\frac{2ev_F}{\hbar}{\cal E}_x + \frac{eNb}{\hbar}\dot{\cal E}_y,
\label{eq:fric}
\end{equation}
where $\tau$ is the relaxation time and $\omega_0$ is the pinning
frequency.  Solving Eq.\ (\ref{eq:fric}) via the Fourier
transformation from the time $t$ to the frequency $\omega$ and
substituting the result into Eqs.\ (\ref{eq:jx}) and (\ref{eq:jy}), we
find the Hall conductivity as a function of frequency:
\begin{equation}
  \sigma_{xy}(\omega)=\frac{2Ne^2}{h}\frac{\omega_0^2-i\omega/\tau}
  {\omega_0^2-\omega^2-i\omega/\tau}.
\label{eq:omega}
\end{equation}
The absolute value of the Hall conductivity, $|\sigma_{xy}|$, computed
from Eq.\ (\ref{eq:omega}) is plotted in Fig.\ \ref{fig:Hall(omega)}
as a function of $\omega/\omega_0$ for $\omega_0\tau=2$.  As we can
see in the figure, the Hall conductivity is quantized at zero
frequency and has a resonance at the pinning frequency.  At the higher
frequencies, where pinning and damping can be neglected and the system
effectively behaves as an ideal, purely inertial system considered in
this section, the Hall conductivity does decrease toward zero.

In this section, the derivation of results was heuristic.  In the
following sections, we calculate the effective action of a moving FISDW
systematically, within the functional-integral formalism.

\section{EFFECTIVE ACTION FOR A (1+1)D DENSITY WAVE} 
\label{sec:(1+1)DW}

As a warmup exercise, let us derive the effective action for a regular
CDW/SDW in the (1+1)D case, where 1+1 represents the space coordinate
$x$ and the time coordinate $t$.  For simplicity, we consider the case
of CDW; results for SDW are the same.  Summation over the spin indices
of electrons is assumed everywhere, which generates a factor of 2 in
traces over the fermions.

Let us consider (1+1)D fermions, described by a Grassmann field
$\Psi(t,x)$, in the presence a density-wave potential
$2\Delta\cos[2k_Fx+\Theta(t,x)]$ and an infinitesimal external
electromagnetic field, described by the scalar $\phi(t,x)$ and vector
$A_x(t,x)$ potentials.  The
action of the system is
\begin{eqnarray}
  &&S[\Psi,\Theta,\phi,A_x]=\int dt\,dx\, \Psi^+
  \biggl[\left(i\hbar\frac{\partial}{\partial t}-e\phi\right)
\label{eq:S_Psi} \\
  &&{}-\frac{1}{2m}
  \left(-i\hbar\frac{\partial}{\partial x}-\frac{e}{c} A_x\right)^2
  +\varepsilon_F-2\Delta\cos(2k_Fx+\Theta) \biggr] \Psi.
\nonumber
\end{eqnarray}

Let us introduce the doublet of fermion fields
\begin{equation}
  \psi(t,x)={\psi_+(t,x) \choose \psi_-(t,x)}
\label{eq:psi}
\end{equation}
with the momenta close to $\pm k_F$:
\begin{equation}
  \Psi(t,x)=\psi_+(t,x)e^{ik_Fx}+\psi_-(t,x)e^{-ik_Fx}.
\label{eq:Psi}
\end{equation}
Substituting Eq.\ (\ref{eq:Psi}) into Eq.\ (\ref{eq:S_Psi}) and
neglecting the terms with the higher derivatives
$\partial^2\psi_\pm/\partial x^2$ and the terms where the
fast-oscillating factors $\exp(\pm i2k_Fx)$ do not cancel out, we
rewrite the action of the system in the matrix form
\begin{equation}
  S[\psi,\Theta,\phi,A_x]={\rm Tr}\int dt\,dx\,
  \psi^+ {\cal L}[\Theta,\phi,A_x] \psi
\label{eq:S_psi}
\end{equation}
with
\begin{eqnarray}
  &&{\cal L}[\Theta,\phi,A_x]=\tau_0
  \left(i\hbar\frac{\partial}{\partial t}-e\phi\right)
\label{eq:L}\\
  &&{}+\tau_z v_F\left(i\hbar\frac{\partial}{\partial x}
  +\frac{e}{c}A_x\right)
  -\tau_x \Delta e^{-i\tau_z \Theta}
  -\tau_0 \frac{e^2}{2mc^2} A_x^2.
\nonumber
\end{eqnarray}
In Eq.\ (\ref{eq:L}), $\tau_x$, $\tau_y$, $\tau_z$, and $\tau_0$ are
the $2\times2$ Pauli matrices and the unit matrix acting on the
doublet of fermion fields (\ref{eq:psi}). In Eq.\ (\ref{eq:S_psi}),
the trace (Tr) is taken over the $\pm$ components of the fermion field
(\ref{eq:psi}) and the implied spin indices of the fermions.

It is convenient to rewrite Eq.\ (\ref{eq:L}) in a pseudorelativistic
notation:
\begin{eqnarray}
  &&{\cal L}[\Theta,A^\mu]=i\hbar v_F\tau_\mu\frac{\partial}{\partial x_\mu}
  -e\frac{v_F}{c}\tau_\mu A^\mu
\label{eq:Dirac}\\
  && -\tau_x \Delta e^{-i\tau_z \Theta}
  -\tau_0 \frac{e^2}{2mc^2} A_x^2,
\nonumber
\end{eqnarray}
where the index $\mu$ takes the values 0 and 1, and summation over
repeated indices is implied.  The contravariant vectors are defined as
follows:
\begin{equation}
  x^\mu=(v_Ft,x),\;
  A^\mu=\left(\frac{c}{v_F}\phi,A_x\right),\;
  \tau^\mu=(\tau_0,\tau_z).
\label{eq:vectors_1+1}
\end{equation}
The covariant vectors are obtained by applying the metric tensor:
$g_{\mu\nu}=g^{\mu\nu}={\rm diag}(1,-1)$.

We wish to find the effective action of the system, ${\Bbb
  S}[\Theta,A^\mu]$, by carrying out the functional integral over the
fermion fields $\psi$ in the partition function with the action
$S[\psi,\Theta,A^\mu]$:
\begin{equation}
  e^{i{\Bbb S}[\Theta,A^\mu]/\hbar}
  =\frac{\int{\cal D}\psi^+\,{\cal D}\psi\,e^{iS[\psi,\Theta,A^\mu]/\hbar}}
  {\int{\cal D}\psi^+\,{\cal D}\psi\,e^{iS[\psi,0,0]/\hbar}}.
\label{eq:S_eff}
\end{equation}
The functional integral (\ref{eq:S_eff}) with action (\ref{eq:S_psi})
is difficult to treat, because the phase $\Theta(x^\mu)$ in Eq.\ 
(\ref{eq:L}) is space-time dependent.  In order to eliminate this
problem, let us change the integration variable $\psi$ to a new
variable $\tilde{\psi}$ via a chiral transformation characterized by a
unitary matrix $U[\Theta(x^\mu)]$ \cite{Brazovskii76}:
\begin{equation}
  \psi(x^\mu)=U[\Theta(x^\mu)]\,\tilde{\psi}(x^\mu)
  =e^{i\tau_z\Theta(x^\mu)/2}\,\tilde{\psi}(x^\mu).
\label{eq:psi_tilde}
\end{equation}
Written in terms of the new field $\tilde{\psi}$, action
(\ref{eq:S_psi}) becomes
\begin{equation}
  \tilde{S}[\tilde{\psi},\Theta,A^\mu]
  ={\rm Tr}\int dt\,dx\,\tilde{\psi}^+\tilde{\cal L}\tilde{\psi},
\label{eq:S_tilde}
\end{equation}
where
\begin{eqnarray}
  \tilde{\cal L}&=&{\cal L}_0+{\cal L}_1+{\cal L}_2,
\label{eq:L012}\\
  {\cal L}_0&=& i\hbar v_F\tau_\mu\frac{\partial}{\partial x_\mu}
  -\tau_x \Delta,
\label{eq:L0}\\
  {\cal L}_1&=& -e\frac{v_F}{c}\tau_\mu B^\mu,
\label{eq:L1'}\\
  {\cal L}_2&=&-\tau_0\frac{e^2}{2mc^2} A_x^2.
\label{eq:L2'}
\end{eqnarray}
In Eq.\ (\ref{eq:L1'}),
\begin{eqnarray}
  B^\mu&=&A^\mu + a^\mu,
\label{eq:B_mu}\\
  a^\mu&=&\frac{\hbar c}{2e}\varepsilon^{\mu\nu}
  \frac{\partial\Theta}{\partial x^\nu},
\label{eq:a_mu}
\end{eqnarray}
where $\varepsilon^{\mu\nu}$ is the antisymmetric tensor with
$\varepsilon^{01}=1$.  The chiral transformation (\ref{eq:psi_tilde})
eliminates the phase factor $\exp({-i\tau_z\Theta})$ of the order
parameter $\Delta$ from Eq.\ (\ref{eq:L}), so that Lagrangian
(\ref{eq:L0}) acquires a simple form.  As a tradeoff, Lagrangian
(\ref{eq:L1'}) subjects fermions to the effective potential
$B^\mu=A^\mu + a^\mu$ (\ref{eq:B_mu}), which combines the original
electromagnetic potentials $A^\mu$ and the gradients of the phase
$\Theta$ (\ref{eq:a_mu}):
\begin{equation}
  a^0=\frac{\hbar c}{2e}\frac{\partial\Theta}{\partial x},\qquad
  a^1=-\frac{\hbar c}{2ev_F}\frac{\partial\Theta}{\partial t}.
\label{eq:a_tx}
\end{equation}

Because the external electromagnetic potentials $A^\mu$ and the
gradients of $\Theta$ are assumed to be small, the effective
potentials $B^\mu$ are also small and can be treated perturbatively.
Changing $\psi$ to $\tilde{\psi}$ and $S$ to $\tilde{S}$ in Eq.\ 
(\ref{eq:S_eff}), we can calculate the effective action ${\Bbb
  S}[\Theta,A^\mu]$ by making a diagrammatic expansion in powers of
$B^\mu$.  Expanding to the first power of Lagrangian (\ref{eq:L1'})
and averaging over the fermions, we obtain the contribution ${\Bbb
  S}_1$ that is nominally of the first order in $B^\mu$.  Expansion to
the second power of (\ref{eq:L1'}) and the first power of
(\ref{eq:L2'}) gives us the contributions ${\Bbb S}_2'$ and ${\Bbb
  S}_2''$ of the second order in $B^\mu$ and $A_x$.  First we
calculate ${\Bbb S}_2={\Bbb S}_2'+{\Bbb S}_2''$ in Sec.\ 
\ref{sec:perturbative} and then obtain ${\Bbb S}_1$ in Sec.\ 
\ref{sec:anomaly}.

\subsection{The second-order terms of the effective action} 
\label{sec:perturbative}

\begin{figure}
\centerline{\psfig{file=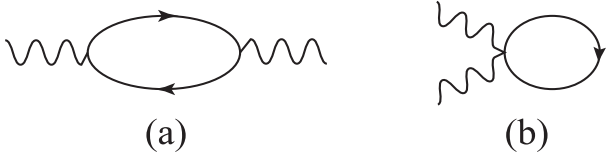,width=\linewidth,angle=90}}
\caption{
  Two Feynman diagrams determining the second-order contribution to
  the effective action, ${\Bbb S}_2$. The solid lines represent the
  fermion Green functions (\ref{eq:G_k}). The wavy lines in panel (a)
  represent the effective potentials $B^\mu$ (\ref{eq:B_mu}), which
  interact with the fermions via Eq.\ (\ref{eq:L1'}).  The wavy lines
  in panel (b) represent the electromagnetic potential $A_x$, which
  interacts with the fermions via Eq.\ (\ref{eq:L2'}) }
\label{fig:diagrams}
\end{figure}

The two second-order contributions to the effective action, ${\Bbb
  S}_2'$ and ${\Bbb S}_2''$, are given by the two Feynman diagrams
shown in Fig.\ \ref{fig:diagrams}, where the wavy lines represent
$B^\mu$ and the solid lines represent the bare Green functions ${\cal
  G}$ of the fermions:
\begin{eqnarray}
  &&{\cal G}(t-t',x-x')=-\frac{i}{\hbar}
  \langle \psi(t,x)\psi^+(x',t') \rangle_{\tilde{S}_0}
\nonumber\\
  &&=\int \frac{dk\,d\omega}{(2\pi)^2} e^{ik(x-x')-i\omega (t-t')} 
  {\cal G}(k,\omega).
\label{eq:G_x}
\end{eqnarray}
The Green function (\ref{eq:G_x}) is obtained by averaging the fermion
fields using action $\tilde{S}_0$ (\ref{eq:S_tilde}) with the
Lagrangian ${\cal L}_0$ (\ref{eq:L0}):
\begin{eqnarray}
  {\cal G}(k,\omega)&=&\frac{e^{i\epsilon\omega}}{\tau_0\hbar\omega
  -\tau_z v_F \hbar k-\tau_x \Delta+i\tau_0\epsilon\,{\rm sgn}(\omega)},
\label{eq:G_k}
\end{eqnarray}
where $\epsilon>0$ is infinitesimal.  Because $\psi$ and $\psi^+$ in
Eq.\ (\ref{eq:G_x}) are two-component fields (\ref{eq:psi}), the Green
function ${\cal G}$ is a $2\times2$ matrix.  The factor
$e^{i\epsilon\omega}$ in Eq.\ (\ref{eq:G_k}) ensures that the integral
in $\omega$ of the Green function (\ref{eq:G_k}),
\begin{equation}
  \int\frac{d\omega}{2\pi}{\rm Tr}[\tau_z {\cal G}(k,\omega)]
  =\frac{2i}{\hbar}\,[n_+(k)-n_-(k)],
\label{eq:n+n-} 
\end{equation}
gives the difference in the occupation numbers $n_+(k)$ and $n_-(k)$
of the $\pm$ fermions. The fermion occupation number $n$ is equal to 1
and 0 at the energies deeply below and high above the Fermi energy,
correspondingly. This statement applies to the electron energies much
greater than the energy gap $\Delta$.  The factor 2 in Eq.\ 
(\ref{eq:n+n-}) comes from the two orientations of the electron spin.

Introducing the Fourier transforms of the potentials
\begin{equation}
  B_\mu(k,\omega)=\int dt\,dx\,
  e^{-ikx + i\omega t}B_\mu(t,x),
\end{equation}
we find an analytical expression for the diagram shown in Fig.\ 
\ref{fig:diagrams}(a)
\begin{equation}
  {\Bbb S}_2'=\frac{e^2v_F^2}{c^2}\int\frac{dp\,d\Omega}{(2\pi)^2} 
  P^{\mu\nu}(p,\Omega) B_\mu(p,\Omega) B_\nu(-p,-\Omega),
\label{eq:S_p1}
\end{equation}
where 
\begin{equation}
  P^{\mu\nu}(p,\Omega)=
  \frac{i\hbar}{2} \int \frac {dk\,d\omega}{(2\pi)^2}
  {\rm Tr}[\tau^\mu{\cal G}(k,\omega)\tau^\nu {\cal G}(k+p,\omega+\Omega)].
\label{eq:P_munu}
\end{equation}
Assuming that the gradients of $B_\mu$ are small, we expand
$P^{\mu\nu}(p,\Omega)$ in powers of $p$ and $\Omega$ and keep only the
zeroth-order term, effectively setting $p=\Omega=0$ in Eq.\ 
(\ref{eq:P_munu}). Thus, we need to calculate the following three
integrals:
\begin{eqnarray}
  &&P^{00}(0,0)=\frac{i\hbar}{2}\int\frac{dk\,d\omega}{(2\pi)^2}
  {\rm Tr}[\tau_0{\cal G}(k,\omega)\tau_0{\cal G}(k,\omega)],
\label{eq:I1} \\
  &&P^{11}(0,0)=\frac{i\hbar}{2}\int\frac{dk\,d\omega}{(2\pi)^2}
  {\rm Tr}[\tau_z{\cal G}(k,\omega)\tau_z{\cal G}(k,\omega)],
\label{eq:I2}\\
  &&P^{10}(0,0)=\frac{i\hbar}{2}\int\frac{dk\,d\omega}{(2\pi)^2}
  {\rm Tr}[\tau_z{\cal G}(k,\omega)\tau_0{\cal G}(k,\omega)].
\label{eq:I3}
\end{eqnarray}
Using Eq.\ (\ref{eq:G_k}) and the identity
\begin{equation}
\partial {\cal G} = -{\cal G}(\partial {\cal G}^{-1}){\cal G},  
\label{eq:dG}
\end{equation}
where $\partial$ represent a derivative of ${\cal G}$ with respect to
any parameter that ${\cal G}$ depends upon, we can rewrite Eqs.\ 
(\ref{eq:I1})--(\ref{eq:I3}) in the following form:
\begin{eqnarray}
  P^{00(10)}&=&-\frac{i}{2} \int \frac{dk\,d\omega}{(2\pi)^2}\,
  {\rm Tr}\left(\tau_{0(z)}
  \frac{\partial{\cal G}(k,\omega)}{\partial\omega}\right),
\label{eq:Idw} \\
  P^{11}&=&\frac{i}{2v_F} \int \frac{dk\,d\omega}{(2\pi)^2} \,
  {\rm Tr}\left(\tau_z\frac{\partial{\cal G}(k,\omega)}{\partial k}\right).
\label{eq:Idk}
\end{eqnarray}
In condensed matter physics, we integrate over the frequency $\omega$
first and than integrate over the wave vector $k$.  Taking the
integral over $\omega$ in Eq.\ (\ref{eq:Idw}), we find that, being an
integral of a full derivative of ${\cal G}(k,\omega)$ with respect to
$\omega$, the integral vanishes, because ${\cal G}(k,\pm\infty)$
vanishes:
\begin{equation}
  P^{00}=P^{10}=0.
\label{eq:I13=0}
\end{equation}
On the other hand, according to Eq.\ (\ref{eq:n+n-}), the
integral over $\omega$ in Eq.\ (\ref{eq:Idk}) gives
\begin{equation}
  P^{11}=-\frac{1}{2\pi\hbar v_F}\int dk\,
  \frac{\partial[n_+(k)-n_-(k)]}{\partial k}=\frac{1}{\pi\hbar v_F}.
\label{eq:I2=}
\end{equation}
We took into account in Eq.\ (\ref{eq:I2=}) that the fermion
occupation number $n$ is equal to 1 and 0 at the energies deeply below
and high above the Fermi energy, correspondingly.

Substituting Eqs.\ (\ref{eq:I13=0}) and (\ref{eq:I2=}) into Eq.\ 
(\ref{eq:S_p1}), we find
\begin{eqnarray}
  {\Bbb S}_2'&=&\frac{e^2v_F}{\pi\hbar c^2}\int dt\,dx\,(B^1)^2
\nonumber \\
  &=&\frac{e^2v_F}{\pi\hbar c^2} \int dt\,dx\,
  \left(A_x
  -\frac{\hbar c}{2ev_F} \frac{\partial\Theta}{\partial t}\right)^2.
\label{eq:S_p1'}
\end{eqnarray}

The analytical expression for the diagram shown in Fig.\ 
\ref{fig:diagrams}(b) is
\begin{equation}
  {\Bbb S}_2''= -\frac{e^2}{2mc^2} \int dt\,dx\,A_x^2\, 
  {\rm Tr}\langle\psi^+(t,x)\tau_0\psi(t,x)\rangle_{\tilde{S}_0}.
\label{eq:S_p2}
\end{equation}
Taking into account that the last factor in Eq.\ (\ref{eq:S_p2}) is
nothing but the average electron density $\rho=4k_F/2\pi$, we find
\begin{equation}
   {\Bbb S}_2''=-\frac{e^2 v_F}{\pi\hbar c^2}\int dt\,dx\,A_x^2.
\label{eq:S_p2'}
\end{equation}

Combining Eqs.\ (\ref{eq:S_p1'}) and (\ref{eq:S_p2'}), we find the
total second-order part of the effective action, ${\Bbb S}_2={\Bbb
  S}_2'+{\Bbb S}_2''$:
\begin{eqnarray}
  &&{\Bbb S}_2[\Theta,A^\mu]=\frac{e^2v_F}{\pi\hbar c^2}
  \int dt\,dx\,[(B^1)^2-(A^1)^2]
\label{eq:S_p} \\
  &&=\int dt\,dx\,
  \biggl[-\frac{e}{c\pi} A_x \frac{\partial\Theta}{\partial t}
  +\frac{\hbar}{4\pi v_F} \left(\frac{\partial\Theta}{\partial t}\right)^2
  \biggr]
\label{eq:S_p''}\\
  &&=\int dt\,dx\,
  \biggl[\frac{e}{\pi c} \Theta \frac{\partial A_x}{\partial t}
  +\frac{\hbar}{4\pi v_F}
  \left(\frac{\partial\Theta}{\partial t}\right)^2\biggr].
\label{eq:S_p'}
\end{eqnarray}
In going from Eq.\ (\ref{eq:S_p''}) to Eq.\ (\ref{eq:S_p'}), we
integrated by parts assuming periodic or zero boundary conditions for
$\Theta$ and $A^i$.  Notice that the $A_x^2$ terms coming from Eqs.\ 
(\ref{eq:S_p1'}) and (\ref{eq:S_p2'}) cancel out exactly, so Eq.\ 
(\ref{eq:S_p'}) does not violate gauge invariance in the absence of
$\Theta$.  When $\Theta\neq0$, it is necessary to add the term ${\Bbb
  S}_1$, calculated in the next section, in order to obtain a
gauge-invariant effective action.

\subsection{The ``first-order'' term of the effective action} 
\label{sec:anomaly}

In the beginning of Sec.\ \ref{sec:(1+1)DW}, we started with a model
(\ref{eq:L}), where the density-wave phase $\Theta(t,x)$ is space-time
dependent.  By doing the chiral transformation (\ref{eq:psi_tilde}) of
the fermions, we made the density-wave phase constant (equal to zero)
in Eq.\ (\ref{eq:L0}) at the expense of modifying the gauge potentials
(\ref{eq:B_mu}).  The chiral transformation
(\ref{eq:psi_tilde}) produces not only a perturbative effect due to
the modification of the gauge potentials, but also changes the ground
state of the system (the ``vacuum'' in the quantum-field-theory
terminology).  Specifically, the chiral transformation changes the
number of fermions in the system, which we calculate below.

Formally, the number of fermions in model (\ref{eq:S_tilde}) is
infinite because of the linearization of the electron dispersion law near
the Fermi energy.  Nevertheless, the {\em variation} of the fermion
number is finite and can be calculated unambiguously, but we need to
introduce some sort of ultraviolet regularization to do this.  When
calculating the fermion density, let us consider the fermion fields at
two points split by a small amount $(\delta x,\delta t)$:
$\rho(t,x)=\langle\psi^+(t+\delta t,x+\delta x)\psi(t,x)\rangle$. The
time splitting is necessary anyway to get the proper time ordering.
Now let us calculate how the fermion number changes when we make an
infinitesimal chiral transformation (\ref{eq:psi_tilde}):
\begin{eqnarray}
  &&\delta\rho(t,x)= 
  \langle\tilde{\psi}^+(t+\delta t,x+\delta x)
\nonumber \\
  &&\{U^+[\delta\Theta(t+\delta t,x+\delta x)]\,
  U[\delta\Theta(t,x)]-1\}\,\tilde{\psi}(t,x)\rangle.
\label{eq:d_rho}
\end{eqnarray}
Expanding the matrices $U$ in $\delta\Theta$ and replacing the average
of the fermions fields by the Green function, we find from Eq.\ 
(\ref{eq:d_rho}):
\begin{eqnarray}
  &&\delta\rho(t,x)=-\frac{\hbar}{2}\,{\rm Tr}\,\{\tau_z
\label{eq:d_rho_G} \\
  &&\times
  [\delta\Theta(t+\delta t,x+\delta x) -
  \delta\Theta(t,x)]\,{\cal G}(-\delta t,-\delta x)\,\}
  \nonumber
\end{eqnarray}
The second line in Eq.\ (\ref{eq:d_rho_G}) can be represented in terms
of the Fourier transforms of $\delta\Theta$ and ${\cal G}$ (see \S19
of Ref.\ \cite{Landau-IX}):
\begin{eqnarray}
  &&\int\frac{dk\,d\omega}{(2\pi)^2}\frac{dp\,d\Omega}{(2\pi)^2}\,
  e^{ipx-i\Omega t-ik\delta x+i\omega\delta t}
\nonumber \\
  &&\times[{\cal G}(k+p,\omega+\Omega)-{\cal G}(k,\omega)]\,
  \delta\Theta(p,\Omega).
\label{eq:G_dxdt}  
\end{eqnarray}
Substituting Eq.\ (\ref{eq:G_dxdt}) into Eq.\ (\ref{eq:d_rho_G}) and
taking the limit $\delta x=\delta t=0$, we find
\begin{eqnarray}
  &&\delta\rho(t,x)=-\frac{\hbar}{2}\int\frac{dp\,d\Omega}{(2\pi)^2}\,
  e^{ipx-i\Omega t}\delta\Theta(p,\Omega)
\nonumber \\
  &&\times{\rm Tr}\,\tau_z\int\frac{dk\,d\omega}{(2\pi)^2}
  [{\cal G}(k+p,\omega+\Omega)-{\cal G}(k,\omega)].
\label{eq:d_rho_q}
\end{eqnarray}
Taking the integral in $\omega$ and the trace as in Eq.\ 
(\ref{eq:n+n-}), we find the following expression for the last line of
Eq.\ (\ref{eq:d_rho_q}):
\begin{eqnarray}
  &&\frac{i}{\hbar}\int\frac{dk}{\pi}
  [n_+(k+p)-n_+(k)-n_-(k+p)+n_-(k)]
\nonumber \\
  &&=\frac{ip}{\pi\hbar}\int dk\,
  \frac{\partial[n_+(k)-n_-(k)]}{\partial k}
  =-\frac{2ip}{\pi\hbar}.
\label{eq:n+-} 
\end{eqnarray}
Ordinarily, by changing the variable of integration $k+p$ to $k$, one
might conclude that integral (\ref{eq:n+-}) vanishes.  However,
because the fermion occupation number $n(k)$ have different values
above and below the Fermi energy, changing the variable of integration
does change the integral, so the result is not zero. To find the
value, we expand Eq.\ (\ref{eq:n+-}) in a series in powers of $p$ and
take the integral over $k$. Only the first term of the series gives a
nonzero result, as shown in the second line of Eq.\ (\ref{eq:n+-}).
Substituting the result into Eq.\ (\ref{eq:d_rho_q}) and performing
the Fourier transform, we find the variation of the fermion density:
\begin{equation}
  \delta\rho(t,x)=\frac{1}{\pi}
  \frac{\partial}{\partial x}\delta\Theta(t,x).
\label{eq:d_rho'} 
\end{equation}
While the local fermion concentration (\ref{eq:d_rho'}) changes, the
total fermion number remains constant:
\begin{equation}
  \int dx\,\delta\rho(t,x)=\frac{1}{\pi}\int dx\,
  \frac{\partial}{\partial x}\delta\Theta(t,x) = 0,
\label{eq:0}
\end{equation}
if we assume that the values of $\Theta(t,x)$ at $x=\pm\infty$ are
equal.  More generally, Eq.\ (\ref{eq:d_rho'}) follows from Eqs.\ 
(\ref{eq:psi}) and (\ref{eq:Psi}), if we notice that a spatial
gradient of $\Theta$ redefines the value of the Fermi momentum $k_F$
and thus changes the number of particles in the Fermi sea.

The variation of the fermion density contributes to the effective
action in the following way.  By averaging Eqs.\ (\ref{eq:S_psi}) and
(\ref{eq:L}) with respect to $\psi$, we find that the electric
potential $\phi$ produces the first-order contribution $-\int
dt\,dx\,e\phi(t,x)\rho(t,x)$ to the effective action.  A chiral
transformation varies the fermion concentration (\ref{eq:d_rho'}), as
well as replaces $\phi$ by the effective potential $B^0$
(\ref{eq:B_mu}). Thus, an infinitesimal chiral transformation results
in the following addition to the effective action:
\begin{eqnarray}
  \delta{\Bbb S}_1&=&-e\frac{v_F}{c}\int
  dt\,dx\,B^0(t,x)\,\delta\rho(t,x)
\nonumber \\
  &=&-\frac{e}{\pi}\frac{v_F}{c}\int dt\,dx\,B^0\,
  \frac{\partial\delta\Theta}{\partial x}.
\label{eq:dS_a}  
\end{eqnarray}
Because the effective potential $B^0$ (\ref{eq:B_mu}) itself depends
on $\Theta$, we need to take a variational integral of Eq.\ 
(\ref{eq:dS_a}) over $\delta\Theta$ in order to recover ${\Bbb S}_1$:
\begin{eqnarray}
  {\Bbb S}_1&=& -\int dt\,dx\,
  \left[\frac{e}{\pi} \phi \frac{\partial\Theta}{\partial x}
  +\frac{\hbar v_F}{4\pi} 
  \left(\frac{\partial\Theta}{\partial x}\right)^2 \right]
\nonumber \\
  &=&\int dx\, dt\,
  \left[\frac{e}{\pi} \Theta \frac{\partial\phi}{\partial x}
  -\frac{\hbar v_F}{4\pi}
  \left(\frac{\partial\Theta}{\partial x}\right)^2 \right].
\label{eq:S_a}
\end{eqnarray}
Action (\ref{eq:S_a}) can be also written in a form similar to Eq.\ 
(\ref{eq:S_p}):
\begin{equation}
  {\Bbb S}_1[\Theta,A^\mu]=
  -\frac{e^2v_F}{\pi\hbar c^2}\int dt\,dx\,[(B^0)^2-(A^0)^2].
\label{eq:S_a'}
\end{equation}
As we see in Eq.\ (\ref{eq:S_a'}), the action ${\Bbb S}_1$ is actually
quadratic in $B^0$, so this action can be called the ``first-order''
term only nominally.  One can easily check explicitly that our
point-splitting method produces zero contribution $\delta j_x$ to
another ``first-order'' term originating from Eq.\ (\ref{eq:L1'}) and
involving $B^1\delta j_x$.

\subsection{The total effective action} 
\label{sec:total}

Equations (\ref{eq:S_p}), (\ref{eq:S_p'}), (\ref{eq:S_a}), and
(\ref{eq:S_a'}) together give the total gauge-invariant effective
action for the (1+1)D density-wave system:
\begin{equation}
  {\Bbb S}[\Theta,A^\mu]={\Bbb S}_2+{\Bbb S}_1
  =\int dt\,dx\,L[\Theta,A^\mu],
\label{eq:S_pa}
\end{equation}
where
\begin{eqnarray}
  L&=&-\frac{e^2v_F}{\pi\hbar c^2}\,[(A_\mu+a_\mu)(A^\mu+a^\mu)-A_\mu A^\mu]
\label{eq:Aa_1+1} \\
  &=&\frac{\hbar}{4\pi v_F}\left(\frac{\partial\Theta}{\partial t}\right)^2 
  -\frac{\hbar v_F}{4\pi}\left(\frac{\partial\Theta}{\partial x}\right)^2
  -\frac{e}{\pi}\Theta {\cal E}_x 
\label{eq:L_eff}
\end{eqnarray}
is the total effective Lagrangian density of the system.  In the
r.h.s.\ of Eq.\ (\ref{eq:L_eff}), the first term represents the
kinetic energy of a rigid displacement of the density wave. The second
term represents the energy change caused by compression or stretching
of the density wave. The third term describes interaction of the
density wave with the electric field. 

Varying $L$ (\ref{eq:L_eff}) with respect to the scalar and
vector potentials, $\phi$ and $A_x$, we find the electric charge
density $\rho_e$ and current density $j_x$ per chain:
\begin{eqnarray}
  \rho_e&=&-\frac{\delta L}{\delta\phi}
  =\frac{e}{\pi}\frac{\partial\Theta}{\partial x},
\label{eq:rho_1+1} \\
  j_x&=&\frac{c\delta L}{\delta A_x}
  =-\frac{e}{\pi}\frac{\partial\Theta}{\partial t}.
\label{eq:jx_1+1}
\end{eqnarray}
Varying Eq.\ (\ref{eq:L_eff}) with respect to $\Theta$, we find the
equation of motion for $\Theta$:
\begin{equation}
  \frac{\partial^2\Theta}{\partial t^2}
  -v_F^2 \frac{\partial^2\Theta}{\partial x^2}
  =-\frac{2ev_F}{\hbar}{\cal E}_x.
\label{eq:EOM_1+1}
\end{equation}

These results are consistent with the standard description of
CDW/SDW \cite{Gruner88}.  Lagrangian (\ref{eq:Aa_1+1}) is often
associated with the so-called (1+1)D chiral anomaly in the quantum
field theory \cite{Krive85,Su86} (see also Ref.\ 
\cite{Brazovskii93}).  Our method of derivation is close to that of
Ref.\ \cite{Ishikawa88}.

\section{EFFECTIVE ACTION FOR (2+1)D FISDW} 
\label{sec:2+1}

Now let us derive the effective action for FISDW, which is (2+1)
dimensional.  We generalize the pseudorelativistic notation
(\ref{eq:vectors_1+1}) to the (2+1)D case as follows:
\begin{eqnarray}
  &&x^i=(v_Ft,x,y),\quad
  A^i=\left(\frac{c}{v_F}\phi,A_x,A_y\right),
\label{eq:vectors_2+1} \\
  &&g_{ij}=g^{ij}={\rm diag}(1,-1,-1).
\nonumber
\end{eqnarray}
We will use roman indices, such as $i$, to denote the (2+1)D vectors
and greek indices, such as $\mu$, to denote the (1+1)D vectors.

It is convenient to Fourier-transform the fields $\psi$, $\phi$, and
$A_x$ over the transverse (discrete) coordinate $y$.  In this
representation, the action of the system is
\begin{eqnarray}
  S[\psi,\Theta,A^i]&=&{\rm Tr}\int 
  \frac{dk_y\,dp_y}{b(2\pi)^2}\,dt\,dx\,\psi^+(t,x,k_y+p_y) 
\nonumber \\
  &&\times {\cal L}[\Theta(t,x,y),A^i(t,x,p_y)]\,\psi(t,x,k_y),
\label{eq:S}
\end{eqnarray}
where $k_y$ and $p_y$ are the wave vectors along the $y$ axis, and
\begin{eqnarray}
  {\cal L}[\Theta,A^i]&=&i\hbar v_F\tau_\mu\frac{\partial}{\partial x_\mu}
  -e\frac{v_F}{c}\tau_\mu A^\mu
\nonumber\\
  &&{}-\tau_x \Delta e^{i\tau_z(NGx-\Theta)}
  -\tau_0 \frac{e^2}{2mc^2} A_x^2
\nonumber\\
  &&{}-\tau_0 2t_b\cos\left(k_y b-Gx-\frac{eb}{\hbar c}A_y\right).
\label{eq:L_2+1}
\end{eqnarray}
The (2+1)D Lagrangian (\ref{eq:L_2+1}) agrees with Eq.\ (\ref{eq:H})
and differs from the (1+1)D Lagrangian (\ref{eq:L}) by the last line
representing the electron tunneling between the chains.  Also, the
FISDW potential has the additional phase $NGx$, because the wave
vector of FISDW is $Q_x=2k_F-NG$, not $2k_F$ as in Sec.\ 
\ref{sec:(1+1)DW}.  The potentials $A^\mu(t,x,p_y)$ in Eqs.\ 
(\ref{eq:S}) and (\ref{eq:L_2+1}) are the Fourier transforms of
$A^\mu(t,x,y)$ over $y$, except for the quadratic term $A_x^2$, which
represents the Fourier transform of the square $A_x^2(t,x,y)$, not the
square of the Fourier transform. We select the gauge $\partial
A_y/\partial y=0$, so $A_y$ does not depend on $y$.  Given that
$\Theta$ may depend on $y$, the factor $\exp(-i\tau_z\Theta)$ in Eq.\ 
(\ref{eq:L_2+1}) symbolically represents the Fourier transform $\int
dy \exp[-ip_yy-i\tau_z\Theta(t,x,y)]$.

\subsection{Transformation of Lagrangian}
\label{sec:transformation}

In this section, we perform two chiral transformations of the fermion
fields that convert the (2+1)D Lagrangian (\ref{eq:L_2+1}) into the
effective (1+1)D form (\ref{eq:L012})--(\ref{eq:L2'}).  Because the
transformations will depend on the transverse wave vector $k_y$, let
us derive a useful formula for such transformations.  Suppose we make
a unitary transformation of the fermion field $\psi(k_y)$, and the
transformation involves a function $f(k_y)$ that depends on $k_y$:
\begin{equation}
  \psi(k_y)=e^{if(k_y)}\tilde{\psi}(k_y).
\label{eq:f_ky}
\end{equation}
Then, a typical term in the Lagrangian transforms in the following
way:
\begin{eqnarray}
  &&\psi^+(k_y+p_y)\phi(p_y)\psi(k_y)
\label{eq:df} \\
  &&\approx
  \tilde{\psi}^+(k_y+p_y)
  \left(1-ip_y\frac{\partial f(k_y)}{\partial k_y}\right)
  \phi(p_y)\tilde{\psi}(k_y).
\nonumber
\end{eqnarray}
Here we substituted Eq.\ (\ref{eq:f_ky}) into Eq.\ (\ref{eq:df}) and
expanded to the first power of the small wave vector $p_y$.

First we make the following transformation of the fermion field $\psi$
in Eqs.\ (\ref{eq:S}) and (\ref{eq:L_2+1}):
\begin{equation}
  \psi=\exp\left[i\tau_z\frac{2t_b}{\hbar v_FG} 
  \sin\left(k_yb-Gx-\frac{eb}{\hbar c}A_y\right)\right]\psi'.
\label{eq:psi'} 
\end{equation}
Written in terms of the fermion field $\psi'$, the Lagrangian of the
system becomes
\begin{eqnarray}
  &&{\cal L}'[\Theta,A^i]\approx 
  i\hbar v_F\tau_\mu\frac{\partial}{\partial x_\mu}
  -e\frac{v_F}{c}\tau_\mu A^\mu
\nonumber\\
  &&-\tau_x\Delta e^{i\tau_z\{NGx-\Theta
  +(4t_b/\hbar v_FG)\sin[k_y b-Gx-(eb/\hbar c)A_y]\}}
\nonumber \\
  &&{}-\tau_0 \frac{e^2}{2mc^2} A_x^2.
\label{eq:L'_2+1}
\end{eqnarray}
As we see in the second line of Eq.\ (\ref{eq:L'_2+1}), transformation
(\ref{eq:psi'}) transfers the interchain hopping term to the FISDW
phase.  Transformation (\ref{eq:psi'}) also generates several terms
proportional to the gradients of $A^i$ and multiplied by the
oscillatory factor $\cos[k_yb-Gx-(eb/\hbar c)A_y]$, which are not
shown in Eq.\ (\ref{eq:L'_2+1}).  These terms would be necessary to
consider if we wanted to keep the terms proportional to $(\partial
A^i/\partial x_j)^2$ in the effective action.  However, since we keep
only the terms with the first derivatives of $A^i$ in the effective
action, the neglected terms are not important, because the oscillatory
factors $\cos(k_yb-Gx)$ would average them to zero.

In the second line of Eq.\ (\ref{eq:L'_2+1}), we expand the interchain
hopping term into the Fourier series:
\begin{equation}
  \exp\left(i\tau_z\frac{4t_b}{\hbar v_FG}\sin\varphi\right)
  =\sum_n J_n\left(\frac{4t_b}{\hbar v_FG}\right) e^{i\tau_z n\varphi},
\label{eq:Bessel}
\end{equation}
where $\varphi=k_yb-Gx-(eb/\hbar c)A_y$, and $J_n(4t_b/\hbar v_FG)$ is
the Bessel function of the integer order $n$ and the argument
$4t_b/\hbar v_FG$.  We neglect all terms except the term with $n=N$ in
series (\ref{eq:Bessel}), because only this term, when substituted
into Eq.\ (\ref{eq:L'_2+1}), does not have oscillatory dependence on
$x$ and opens an energy gap at the Fermi level. This is the so-called
single-gap approximation, well-known in the theory of FISDW
\cite{Montambaux86,Maki86b,Yakovenko91a}. In this way we obtain the
following approximate expression for Lagrangian (\ref{eq:L'_2+1}):
\begin{eqnarray}
  &&{\cal L}'[\Theta,A^i]\approx 
  i\hbar v_F\tau_\mu\frac{\partial}{\partial x_\mu}
  -e\frac{v_F}{c}\tau_\mu A^\mu
\nonumber \\
  &&{}-\tau_0 \frac{e^2}{2mc^2} A_x^2
  -\tau_x\tilde{\Delta} 
  e^{i\tau_z\{Nb[k_y-(e/\hbar c)A_y]-\Theta\}}, 
\label{eq:L'_approx}
\end{eqnarray}
where $\tilde{\Delta}=\Delta J_N(4t_b/\hbar v_FG)$. The transformed
Lagrangian (\ref{eq:L'_approx}) of the (2+1)D FISDW is the same as
Lagrangian (\ref{eq:L}) of the (1+1)D density wave with the
replacement $\Delta\to\tilde\Delta$ and $\Theta\to\tilde\Theta$, where
\begin{equation}
  \tilde\Theta=\Theta+Nb[(e/\hbar c)A_y-k_y].
\label{eq:Theta_tilde}  
\end{equation}

Now we make the second transformation of the fermions \cite{Theta_y}:
\begin{equation}
  \psi'=\exp\left\{i\frac{\tau_z}{2}\left[\Theta
  -Nb\left(k_y-\frac{e}{\hbar c}A_y\right)\right] \right\} 
  \tilde{\psi}.
\label{eq:psi_tilde_2+1}
\end{equation}
This chiral transformation eliminates the phase of the FISDW potential
in the last term of Eq.\ (\ref{eq:L'_approx}), and the transformed
action becomes:
\begin{equation}
  \tilde{S}[\tilde{\psi},\Theta,A^\mu]=\frac{1}{b}\,{\rm Tr}
  \int dt\,dx\,dy\,
  \tilde{\psi}^+\tilde{\cal L}\tilde{\psi}, 
\label{eq:S_tilde_2+1}
\end{equation}
where $\tilde{\cal L}$ has the (1+1)D form
(\ref{eq:L012})--(\ref{eq:L2'}) with $\Delta\to\tilde{\Delta}$ and the
new effective potentials
\begin{eqnarray}
  \tilde B^\mu&=&A^\mu+a^\mu-\frac{Nb}{2}
  \varepsilon^{\mu ij} \frac{\partial A_j}{\partial x^i},
\label{eq:B_mu_tilde} \\
  \tilde B^0&=&\frac{c}{v_F}\phi
  +\frac{\hbar c}{2e}\frac{\partial\Theta}{\partial x}
  +\frac{Nb}{2}{\cal H}_z,
\label{eq:A_t_2+1} \\
  \tilde B^1 &=&A_x-\frac{\hbar c}{2ev_F}
  \frac{\partial\Theta}{\partial t}+\frac{Nbc}{2v_F}{\cal E}_y.
\label{eq:A_x_2+1}
\end{eqnarray}
The potentials (\ref{eq:A_t_2+1}) and (\ref{eq:A_x_2+1}) differ from
the corresponding (1+1)D expressions (\ref{eq:B_mu}) by the extra
terms proportional to the integer $N$ and the electromagnetic fields
${\cal H}_z=\partial A_y/\partial x-\partial A_x/\partial y$ and
${\cal E}_y=-\partial A_y/c\partial t-\partial \phi/\partial y$.  The
terms $\partial A_y/\partial x$ and $\partial A_y/\partial t$ appear
in Eqs.\ (\ref{eq:A_t_2+1}) and (\ref{eq:A_x_2+1}) when the
differential operator in Eq.\ (\ref{eq:L'_approx}) is applied to
transformation (\ref{eq:psi_tilde_2+1}).  The terms
$\partial\phi/\partial y$ and $\partial A_x/\partial y$ appear when we
apply Eq.\ (\ref{eq:df}) to transformation (\ref{eq:psi_tilde_2+1})
with $f(k_y)=-Nbk_y\tau_z/2$ and convert $ip_y A^\mu(p_y)$ into
$\partial A^\mu/\partial y$.  It was possible to reduce the (2+1)D
Lagrangian (\ref{eq:L_2+1}) into the effectively (1+1)D one only
because the magnetic field $H$ suppressed the fermion energy
dispersion in $k_y$, which made the system effectively (1+1)D.

\subsection{Effective action}
\label{sec:effective}

The ``second-order'' part of the effective action for FISDW is
obtained immediately by substituting Eq.\ (\ref{eq:A_x_2+1}) into Eq.\ 
(\ref{eq:S_p}):
\begin{eqnarray}
  {\Bbb S}_2&=&\frac{e^2v_Fb}{\pi\hbar c^2}
  \int dt\,dx\,dy\,[(\tilde B^1)^2-(A^1)^2]
\label{eq:S_p_2+1'} \\
  &=&\int dt\,dx\,dy\,\biggl[
  \frac{\hbar}{4\pi v_Fb}\left(\frac{\partial\Theta}{\partial t}\right)^2 
  +\frac{e}{\pi bc}\Theta\frac{\partial A_x}{\partial t}
\nonumber \\
  &&{}-\frac{Ne}{2\pi v_F} {\cal E}_y\frac{\partial\Theta}{\partial t}
  +\frac{Ne^2}{\pi\hbar c}A_x{\cal E}_y\biggr].
\label{eq:S_p_2+1}
\end{eqnarray}
We neglected the term proportional to ${\cal E}_y^2$ in Eq.\ 
(\ref{eq:S_p_2+1}).

To determine the ``first-order'' part of the effective action, we need
to find the variation of the fermion density associated with
transformations (\ref{eq:psi'}) and (\ref{eq:psi_tilde_2+1}) following
the method of Sec.\ \ref{sec:anomaly}.  We neglect the contribution
from the first transformation (\ref{eq:psi'}) because of the
oscillatory factor $\cos(k_yb-Gx)$.  From Eq.\ (\ref{eq:dS_a}) we find
that the second transformation (\ref{eq:psi_tilde_2+1}) gives the
following contribution to the effective action:
\begin{equation}
  \delta{\Bbb S}_1 = \frac{1}{\pi b}\int dt\,dx\,dy\,\tilde B_t
  \left(\frac{\partial\delta\Theta}{\partial x}
  +\frac{Neb}{\hbar c}\frac{\partial\delta A_y}{\partial x}\right).
\label{eq:dS_a_2+1}  
\end{equation}
Because transformation (\ref{eq:psi_tilde_2+1}) depends on the two
parameters $\Theta$ and $A_y$, Eq.\ (\ref{eq:dS_a_2+1}) contains the
variations of both.  Substituting $\tilde B_t$ from Eq.\ 
(\ref{eq:A_t_2+1}) into Eq.\ (\ref{eq:dS_a_2+1}) and taking the
variational integral over $\delta\Theta$ and $\delta A_y$, we get the
``first-order'' part of the action:
\begin{eqnarray}
  {\Bbb S}_1&=&\int dt\,dx\,dy\,
  \biggl[\frac{e}{\pi b} \Theta \frac{\partial\phi}{\partial x}
  -\frac{\hbar v_F}{4\pi b}
  \left(\frac{\partial\Theta}{\partial x}\right)^2 
\nonumber\\
  &&{}-\frac{Ne v_F}{2 \pi c} {\cal H}_z \frac{\partial\Theta}{\partial x}
  +\frac{Ne^2}{\pi\hbar c}A_y\frac{\partial\phi}{\partial x}\biggr],
\label{eq:S_a_2+1}
\end{eqnarray}
where we neglected the term proportional to ${\cal H}_z\partial A_y/\partial
x$.  Eq.\ (\ref{eq:S_a_2+1}) can be written in the form
\begin{equation}
  {\Bbb S}_1=
  -\frac{e^2v_F}{\pi\hbar c^2b}\int dt\,dx\,dy\,
  \left((\tilde B^0)^2-(A^0)^2
  +\frac{Nbc}{v_F}\phi\frac{\partial A_x}{\partial y}\right)
\label{eq:S_a'_2+1}
\end{equation}
with $\tilde B^0$ given by Eq.\ (\ref{eq:A_t_2+1}).  Eq.\ 
(\ref{eq:S_a'_2+1}) is similar to the (1+1)D Eq.\ (\ref{eq:S_a'}), but
contains the extra last term.

Equations (\ref{eq:S_p_2+1}) and (\ref{eq:S_a_2+1}) together give the
total gauge-invariant effective action of FISDW:
\begin{equation}
  {\Bbb S}[\Theta,A^\mu]={\Bbb S}_2+{\Bbb S}_1
  =\int dt\,dx\,dy\,L[\Theta,A^\mu],
\label{eq:S_pa_2+1}
\end{equation}
where \cite{E_x^2}
\begin{eqnarray}
  L&=&
  \frac{\hbar}{4\pi v_Fb}\left(\frac{\partial\Theta}{\partial t}\right)^2 
  -\frac{\hbar v_F}{4\pi b}\left(\frac{\partial\Theta}{\partial x}\right)^2
  -\frac{e}{\pi b} \Theta {\cal E}_x 
\nonumber\\
  &&{}-\frac{Ne^2}{2\pi\hbar c}(\phi {\cal H}_z-A_x {\cal E}_y+A_y {\cal E}_x)
\nonumber\\
  &&{}-\frac{Ne}{2\pi v_F} {\cal E}_y\frac{\partial\Theta}{\partial t} 
  -\frac{Nev_F}{2\pi c} {\cal H}_z \frac{\partial\Theta}{\partial x}.
\label{eq:L_eff_2+1}
\end{eqnarray}
In Eq.\ (\ref{eq:L_eff_2+1}), the first line is the same as the
Lagrangian density of a purely (1+1)D density wave (\ref{eq:L_eff})
(save for the overall factor $1/b$) and, unlike the next two lines, is
not specific to FISDW.  The second line represents the Chern-Simons
term responsible for QHE in the FISDW state.  The last line describes
the interaction of the FISDW motion and compression with the
transverse electric field ${\cal E}_y$ and the magnetic field ${\cal H}_z$.

When $\Theta$ does not depend on the coordinate $x$, the effective
Lagrangian (\ref{eq:L_eff_2+1}) coincides with the Lagrangian
$L=L_1+L_2$, derived semiphenomenologically in Sec.\ 
\ref{sec:phenomenological} (Eqs.\ (\ref{eq:L1}) and (\ref{eq:L2})).
When the FISDW is pinned and immobile, so that $\Theta$ is not a
dynamical variable, Eq.\ (\ref{eq:L_eff_2+1}) reduces to only the
Chern-Simons term:
\begin{equation}
  L_{\rm CS}=\frac{Ne^2}{2\pi\hbar c}\varepsilon_{ijk}
  A^i\frac{\partial A^k}{\partial x_j}.
\label{eq:CS}
\end{equation}
We can reintroduce the dynamics of FISDW by replacing the
electromagnetic potentials $A^i$ in Eq.\ (\ref{eq:CS}) with the
effective potentials $A^i+a^i$, where $a^\mu$ are given by Eqs.\
(\ref{eq:a_mu}) and (\ref{eq:a_tx}), and the third component is zero:
$a^2=0$.  Adding also the Lagrangian (\ref{eq:Aa_1+1}) of the (1+1)D
density wave, we recover the Lagrangian (\ref{eq:L_eff_2+1}) of the
(2+1)D FISDW in the following form:
\begin{eqnarray}
  &&L=\frac{Ne^2}{2\pi\hbar c}\varepsilon_{ijk}
  (A^i+a^i)\frac{\partial (A^k+a^k)}{\partial x_j}
\nonumber \\
  &&{}-\frac{e^2v_F}{\pi \hbar c^2b}\,
  [(A_\mu+a_\mu)(A^\mu+a^\mu)-A_\mu A^\mu].
\label{eq:CS'}
\end{eqnarray}
Thus, the effective action of FISDW is given simply by the (2+1)D
Chern-Simons term and the (1+1)D chiral anomaly written for the
combined electromagnetic potentials and the FISDW phase gradients
$A^i+a^i$.  The effective Lagrangian of FISDW (\ref{eq:L_eff_2+1}) can
be also written in a (1+1)D form resembling Eq.\ (\ref{eq:Aa_1+1}):
\begin{equation}
  L=-\frac{e^2v_F}{\pi\hbar c^2b}
  \left(\tilde B_\mu\tilde B^\mu-A_\mu A^\mu
  +\frac{Nbc}{v_F}\phi\frac{\partial A_x}{\partial y}\right),
\label{eq:L_eff_2+1'}
\end{equation}
where the effective potentials $\tilde B^\mu$ are given by Eq.\ 
(\ref{eq:B_mu_tilde}).

By varying Eq.\ (\ref{eq:L_eff_2+1}) with respect to $\phi$, $A_x$,
and $A_y$, we find the electric charge density $\rho_e$ and the
current densities $j_x$ along the chains and $j_y$ perpendicular to
the chains:
\begin{eqnarray}
  \rho_e&=& \frac{e}{\pi b}\frac{\partial \Theta}{\partial x}
  +\frac{Ne}{2\pi v_F}\frac{\partial^2\Theta}{\partial y \partial t} 
  +\frac{2 Ne^2}{h c} {\cal H}_z,
\label{eq:rho_2+1} \\
  j_x&=&-\frac{e}{\pi b}\frac{\partial \Theta}{\partial t}
  -\frac{Nev_F}{2\pi}\frac{\partial^2\Theta}{\partial y \partial x}
  +\frac{2 Ne^2}{h } {\cal E}_y,
\label{eq:jx_2+1}\\
  j_y&=&- \frac{2 Ne^2}{h } {\cal E}_x
  -\frac{Ne }{2 \pi v_F} \frac{\partial^2 \Theta}{\partial t^2} 
  +\frac{Ne v_F}{2 \pi } \frac{\partial^2 \Theta}{\partial x^2}.
\label{eq:jy_2+1}
\end{eqnarray}
The equation of motion of $\Theta$ is obtained by varying Eq.\ 
(\ref{eq:L_eff_2+1}) with respect to $\Theta$:
\begin{equation}
  \frac{\partial^2 \Theta}{\partial t^2}-v_F^2
  \frac{\partial^2 \Theta}{\partial x^2}
  =-\frac{2ev_F}{\hbar}{\cal E}_x+\frac{eNb}{\hbar}
  \frac{\partial {\cal E}_y}{\partial t}
  +\frac{eN v_F^2 b}{\hbar c} \frac{\partial {\cal H}_z}{\partial x}.
\label{eq:EOM_2+1}
\end{equation}
When $\Theta$ does not depend on the coordinates $x$ and $y$, Eqs.\ 
(\ref{eq:jx_2+1})--(\ref{eq:EOM_2+1}) coincide with the corresponding
Eqs.\ (\ref{eq:jx}), (\ref{eq:jy}), and (\ref{eq:EOM}) derived
semiphenomenologically in Sec.\ \ref{sec:phenomenological}.  When
$N=0$, Eqs.\ (\ref{eq:rho_2+1}), (\ref{eq:jx_2+1}), and
(\ref{eq:EOM_2+1}) coincide with the corresponding Eqs.\ 
(\ref{eq:rho_1+1})--(\ref{eq:EOM_1+1}) for the (1+1)D density wave.

\section{ALTERNATIVE DERIVATION OF THE EFFECTIVE ACTION}
\label{sec:alternative}

In this section, we briefly outline an alternative derivation of the
effective action for the considered systems.  This derivation will be
straightforwardly generalized to finite temperatures in the next
section.  

We noticed in Sec.\ \ref{sec:2+1} that after transformation
(\ref{eq:psi'}) Lagrangian (\ref{eq:L'_approx}) of the (2+1)D FISDW is
the same as Lagrangian (\ref{eq:L}) of the (1+1)D density wave with
the replacement $\Delta\to\tilde\Delta$ and $\Theta\to\tilde\Theta$,
where $\tilde\Theta$ is given by Eq.\ (\ref{eq:Theta_tilde}).  While,
in principle, the chiral transformation (\ref{eq:psi'}) may bring some
contribution to the effective action of FISDW, this contribution is
not essential in practice, because of the oscillatory factor
$\cos(k_yb-Gx)$, as we observed in Sec.\ \ref{sec:2+1}.  Thus, to find
the effective action for FISDW, as well as for a (1+1)D density wave,
it is sufficient to calculate the effective action for Lagrangian
(\ref{eq:L'_approx}).

Instead of calculating the effective action ${\Bbb S}$ directly, let
us calculate its variation with respect to a variation
$\delta\tilde\Theta$ of phase (\ref{eq:Theta_tilde}):
\begin{equation}
  \delta{\Bbb S}
  =\frac{\int{\cal D}\psi'^+\,{\cal D}\psi'\,\psi'^+\,
  \frac{\delta{\cal L}'}{\delta\tilde\Theta}\delta\tilde\Theta\,
  \psi'\,e^{iS'[\psi',\tilde\Theta]/\hbar}}
  {\int{\cal D}\psi'^+\,{\cal D}\psi'\,e^{iS'[\psi',\tilde\Theta]/\hbar}}.
\label{eq:dS_eff}
\end{equation}
The advantage of Eq.\ (\ref{eq:dS_eff}) is that, after we make
transformation (\ref{eq:psi_tilde_2+1}), the anomalous terms cancel
out in numerator and denominator, so it is sufficient to calculate
only a perturbative contribution to $\delta{\Bbb S}$.  Using the
explicit form of ${\cal L}'$ (\ref{eq:L'_approx}) and taking the
variation in Eq.\ (\ref{eq:dS_eff}), we find:
\begin{equation}
  \frac{\delta{\Bbb S}}{\delta\tilde\Theta}
  =\langle\psi'^+\tilde\Delta\tau_ye^{-i\tau_z\tilde\Theta}\psi'\rangle_{S'}
  =\langle\tilde\psi^+\tilde\Delta\tau_y 
  e^{-i\tau_z\Theta_0}\tilde\psi\rangle_{\tilde{S}}\;.
\label{eq:tau2}
\end{equation}
Instead of transformation (\ref{eq:psi_tilde_2+1}), we made a slightly
different transformation
\begin{eqnarray}
  \psi'&=&e^{i\tau_z(\tilde\Theta-\Theta_0)/2}\tilde{\psi},
\label{eq:psi_tilde'}
\end{eqnarray}
which changes the density-wave phase not to zero, but to a constant,
space-time-independent value $\Theta_0$.

In Eq.\ (\ref{eq:tau2}), we expand $\tilde{S}$, given by Eqs.\ 
(\ref{eq:S_tilde_2+1}), to the first order in $\tilde B^\mu$ and find
the following expression in the momentum representation:
\begin{eqnarray}
  \delta{\Bbb S}=-e\frac{v_F}{c}\int\frac{dp_x\,dp_y\,d\Omega}{(2\pi)^3}
  && \Pi^\mu(p_x,\Omega) \tilde B_\mu(-p_x,-p_y,-\Omega)
\nonumber \\
  &&\times\delta\tilde\Theta(p_x,p_y,\Omega),
\label{eq:dS_Pi}
\end{eqnarray}
where
\begin{eqnarray}
  \Pi^\mu&=&\frac{i\hbar\tilde\Delta}{b} 
  \int\frac{dk\,d\omega}{(2\pi)^2}
\label{eq:Pi_mu} \\
  &&\times{\rm Tr}\,[\tau_y e^{-i\tau_z\Theta_0}{\cal G}(k,\omega)
  \tau^\mu{\cal G}(k+p,\omega+\Omega)]
\nonumber
\end{eqnarray}
with ${\cal G}$ defined by Eq.\ (\ref{eq:G_k}) with
$\Delta\to\tilde\Delta\exp(-i\tau_z\Theta_0)$.  Expanding Eq.\ 
(\ref{eq:Pi_mu}) to the first order in $p$ and $\Omega$, we rewrite
Eq.\ (\ref{eq:dS_Pi}) as follows:
\begin{eqnarray}
  \delta{\Bbb S}=e\frac{v_F}{c}\int\frac{dp_x\,dp_y\,d\Omega}{(2\pi)^3}
  && Q^{\mu\nu}\frac{ip_\nu}{\pi b} 
  \tilde B_\mu(-p_x,-p_y,-\Omega)
\nonumber \\
  &&\times\delta\tilde\Theta(p_x,p_y,\Omega),
\label{eq:dS_Q}
\end{eqnarray}
where $p_0=\Omega/v_F$, $p_1=-p_x$, and 
\begin{eqnarray}
  && Q^{\mu\nu}=\tilde\Delta\pi\hbar^2v_F\int\frac{dk\,d\omega}{(2\pi)^2}
\label{eq:Q_munu} \\
  && \times{\rm Tr}\,[\tau_ye^{-i\tau_z\Theta_0}{\cal G}(k,\omega)
  \tau^\mu{\cal G}(k,\omega)
  \tau^\nu{\cal G}(k,\omega)].
\nonumber
\end{eqnarray}
Manipulating the $\tau$ matrices in Eq.\ (\ref{eq:Q_munu}), it is
possible to show that the tensor $Q^{\mu\nu}$ is antisymmetric:
\begin{equation}
  Q^{\mu\nu}=-\frac12C\varepsilon^{\mu\nu}.
\label{eq:QC}
\end{equation}
The constant $C$ in Eq.\ (\ref{eq:QC}) is an integer topological
invariant, the Chern number:
\begin{equation}
  C=\int\frac{dk\,d\omega\,d\Theta_0}{4\pi^2}
  {\rm Tr}\left(
  \frac{\partial {\cal G}^{-1}}{\partial\Theta_0} {\cal G}
  \frac{\partial {\cal G}^{-1}}{\partial\omega} {\cal G}
  \frac{\partial {\cal G}^{-1}}{\partial k} {\cal G}\right)=-2.
\label{eq:C}
\end{equation}
In Eq.\ (\ref{eq:C}), the fermion Green function ${\cal G}$
(\ref{eq:G_k}) with $\Delta\to\tilde\Delta\exp(-i\tau_z\Theta_0)$ is a
function of three variables: $\omega$, $k$, and the density-wave phase
$\Theta_0$.  The integral over $\Theta_0$ has been added in Eq.\
(\ref{eq:C}), because the result does not depend on the value of
$\Theta_0$.  Integral (\ref{eq:C}) is calculated in the Appendix.  The
value 2 comes from the two orientations of the electron spin.
Substituting Eqs.\ (\ref{eq:QC}) and (\ref{eq:C}) into Eq.\
(\ref{eq:dS_Q}) and Fourier-transforming to the real space, we find:
\begin{equation}
  \delta{\Bbb S}=-\frac{ev_F}{\pi cb}\int dt\,dx\,dy\,
  \varepsilon^{\mu\nu}\,\tilde B_\mu(x,y,t)\,
  \frac{\partial\delta\tilde\Theta(x,y,t)}{\partial x^\nu}.
\label{eq:dS_BW}
\end{equation}
Taking the variational integral over $\delta\tilde\Theta$ in Eq.\ 
(\ref{eq:dS_BW}), we recover the effective action
(\ref{eq:L_eff_2+1}).

\section{TEMPERATURE DEPENDENCE OF THE HALL EFFECT} 
\label{sec:temperature}

The Hall conductivity at a finite temperature is not quantized because
of the presence of thermally excited quasiparticles above the energy
gap. It is interesting to find how the Hall conductivity evolves with
the temperature. Because QHE at zero temperature is generated by the
collective motion of electrons in the FISDW condensate, the issue here
is the temperature dependence of the condensate current.  One would
expect that the condensate current must gradually decrease with
increasing temperature and vanish at the transition temperature $T_c$,
where the FISDW order parameter disappears.  This behavior is
qualitatively similar to the temperature evolution of the
superconducting condensate density and the inverse penetration depth
of magnetic field in superconductors.

We start our consideration from the transformed Lagrangian
(\ref{eq:L'_approx}) of the (2+1)D FISDW, which is the same as
Lagrangian (\ref{eq:L}) of the (1+1)D density wave with the effective
phase $\tilde\Theta$ (\ref{eq:Theta_tilde}) instead of $\Theta$.  A
time dependence of $\tilde\Theta$ generates the Fr\"{o}hlich current
along the chains:
\begin{equation}
  j_x=-\frac{e}{\pi b}\frac{\partial\tilde\Theta}{\partial t}.
\label{jxphi}
\end{equation}
In the presence of a transverse electric field ${\cal E}_y$, we have
$A_y=-c{\cal E}_yt$ in Eq.\ (\ref{eq:Theta_tilde}), then Eq.\ 
(\ref{jxphi}) reproduces Eq.\ (\ref{eq:jx}).  If FISDW is pinned
($\dot{\Theta}=0$), then Eq.\ (\ref{jxphi}) describes QHE.  So, the
quantum Hall conductivity is the Fr\"{o}hlich conductivity associated
with the combined phase $\tilde\Theta$ (\ref{eq:Theta_tilde}).  Thus,
the temperature dependence of QHE must be the same as the temperature
dependence of the Fr\"ohlich conductivity. The latter issue was
studied in the theory of a regular CDW/SDW \cite{Lee79,Maki90}. It was
found that, at a finite temperature $T$, the Fr\"ohlich current
carried by the CDW/SDW condensate is reduced with respect to the
zero-temperature value (\ref{jxphi}) by a factor $f(T)$:
\begin{equation}
  j_x=-f(T)\,\frac{e}{\pi b}\frac{\partial\tilde\Theta}{\partial t}.
\label{f(T)jxphi}
\end{equation}
We conclude that the same factor $f(T)$ reduces the Hall conductivity
of a pinned FISDW:
\begin{equation}
  \sigma_{xy}(T)=f(T)\,\frac{2Ne^2}{h}.
\label{f(T)2Ne2/h}
\end{equation}
In Eqs.\ (\ref{f(T)jxphi}) and (\ref{f(T)2Ne2/h}), the function $f(T)$
is
\begin{equation}
  f(T)=1-\int_{-\infty}^\infty \frac{dk}{\hbar v_F}
  \left(\frac{\partial E_k}{\partial k}\right)^{\!2}
  \left(-\frac{\partial n(E_k)}{\partial E_k}\right),
\label{f(T)}
\end{equation}
where $k_x$ is relabeled as $k$, $E_k=\sqrt{(\hbar
  v_Fk)^2+\tilde\Delta^2}$ is the electron dispersion law in the FISDW
state, and $n(\epsilon)=(e^{\epsilon/T}+1)^{-1}$ is the Fermi
distribution function.  At a finite temperature, normal quasiparticles
thermally excited above the energy gap equilibrate with the immobile
crystal lattice.  Thus, only a fraction of all electrons is carried
along the chains by the moving periodic potential, which reduces the
Hall/Fr\"ohlich current by the last term in Eq.\ (\ref{f(T)}).

\begin{figure}
\centerline{\psfig{file=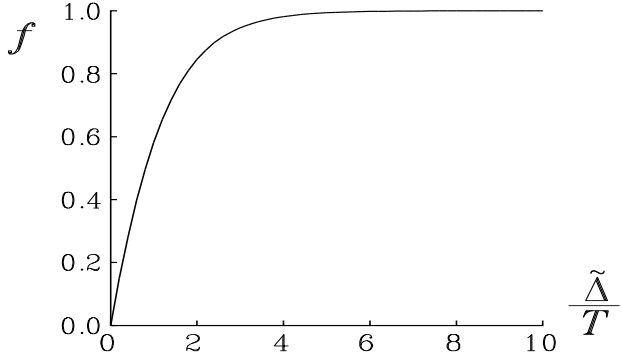,width=\columnwidth,height=0.6\columnwidth,angle=-90}}
\caption{The reduction factor $f$ of the Hall conductivity, given by Eq.\
  (\protect\ref{dynMaki}) and shown as a function of the ratio of the
  energy gap at the Fermi level $\tilde\Delta$ to temperature $T$.}
\label{Fig:f(T)}
\end{figure}

\begin{figure}
\centerline{\psfig{file=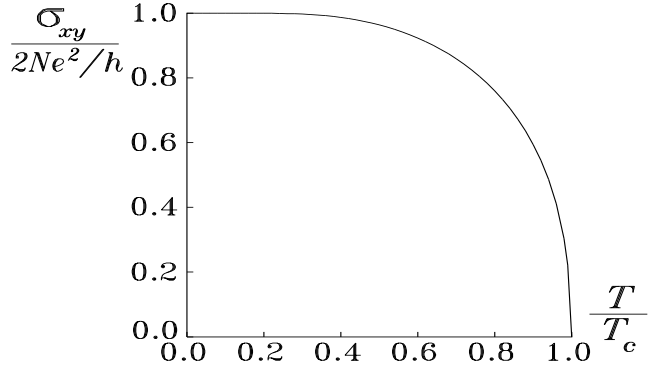,width=\columnwidth,height=0.6\columnwidth,angle=-90}}
\caption{
  Hall conductivity in the FISDW state, $\sigma_{xy}$, as a function
  of temperature $T$ normalized to the FISDW transition temperature
  $T_c$.}
\label{Fig:sxy(T)}
\end{figure}

The function $f$ (\ref{f(T)}) depends only on the ratio of the energy
gap at the Fermi level, $\tilde\Delta$, and the temperature $T$.
Introducing the new variable of integration $\zeta$ instead of $k$ via
the equation $\hbar v_Fk=\tilde{\Delta}\sinh\zeta$, we can rewrite
Eq.\ (\ref{f(T)}) as follows \cite{Maki89,Maki90}:
\begin{equation}
  f\left(\frac{\tilde\Delta}{T}\right)=
  \int_0^\infty d\zeta\,
  \frac{\tanh\left(\frac{\tilde\Delta}{2T}\cosh\zeta\right)}
  {\cosh^2\zeta}.
\label{dynMaki}
\end{equation}
The function $f(\tilde\Delta/T)$ is plotted in Fig.\ 
\ref{Fig:f(T)}. It is equal to 1 at zero temperature, where Eq.\ 
(\ref{f(T)2Ne2/h}) gives QHE, gradually decreases with increasing $T$,
and vanishes when $T\gg\tilde\Delta$. Taking into account that the
FISDW order parameter $\Delta$ itself depends on $T$ and vanishes at
the FISDW transition temperature $T_c$, it is clear that $f(T)$ and
$\sigma_{xy}(T)$ vanish at $T\rightarrow T_c$, where
$\sigma_{xy}(T)\propto f(T)\propto\Delta(T)\propto\sqrt{T_c-T}$.
Assuming that the temperature dependence $\tilde\Delta(T)$ is given by
the BCS theory \cite{Montambaux86}, we plot the temperature dependence
of the Hall conductivity, $\sigma_{xy}(T)$, in Fig.\ \ref{Fig:sxy(T)}
\cite{normal}.

The function $f(T)$ (\ref{f(T)}) is qualitatively similar to the
function $f_{\rm s}(T)$ that describes the temperature reduction of
the superconducting condensate density in the London case. Both
functions approach 1 at zero temperature, but near $T_c$ the
superconducting function behaves differently: $f_{\rm s}(T)\propto
\Delta^2(T)\propto T_c-T$. As explained in Sec.\ \ref{sec:Matsubara},
this is due to the difference between the static and dynamic limits of
the response function.

In the next Sec.\ \ref{sec:Galileo}, we give a simple,
semiphenomenological derivation of Eqs.\ (\ref{f(T)jxphi}) and
(\ref{f(T)}) based on the ideas of Refs.\ \cite{Lee79,Overhauser} and
analogous to the standard derivation of the superfluid density (see
\S27 of Ref.\ \cite{Landau-IX}).  After that, in Sec.\ 
\ref{sec:Matsubara}, we give a formal diagrammatic derivation Eqs.\ 
(\ref{f(T)jxphi}) and (\ref{f(T)}).  We also derive the effective
action of FISDW at a finite temperature.

\subsection{Semiphenomenological derivation}
\label{sec:Galileo}

Let us consider a 1D electron system where a CDW/SDW of an amplitude
$\Delta$ moves with a small velocity $v$. Let us calculate the
Fr\"ohlich current, proportional to $v$, at a finite temperature $T$.

We find the electron wave functions in the reference frame moving with
the density wave and then Galileo-transform them to the laboratory
frame \cite{Overhauser}:
\begin{eqnarray}
  \psi_k^\pm(t,x)&=&u_k^{\pm}
  e^{i(k_{\rm F}+k+mv)x-i(k_{\rm F}+k)vt\mp iE_kt/\hbar}
\nonumber \\
  &&+w_k^{\pm}e^{i(-k_{\rm F}+k+mv)x-i(-k_{\rm F}+k)vt\mp iE_kt/\hbar},
\label{psi} 
\end{eqnarray}
where we keep only the terms linear in $v$. In Eq.\ (\ref{psi}) and
below, the index $\pm$ refers to the states above and below the
CDW/SDW energy gap, {\em not} to the states near $\pm k_{\rm F}$. The
coefficients of superposition $u_k$ and $w_k$ are given by the
following expressions:
\begin{equation}
  |u_k^+|^2=|w_k^-|^2=\frac{\Delta^2}{2E_k(E_k-\xi_k)},
\label{uw1}
\end{equation}
\begin{equation}
  |w_k^+|^2=|u_k^-|^2=\frac{E_k-\xi_k}{2E_k},
\label{uw2} 
\end{equation}
where $\xi_k=\hbar v_Fk$ and $E_k=\sqrt{\xi_k^2+\Delta^2}$ are
the electron dispersion laws in the absence and in the presence of the
CDW/SDW energy gap.

By analogy with the standard derivation of the superfluid density
(\S27 of Ref.\ \cite{Landau-IX}), let us assume that, because of
interaction with impurities, phonons, etc., the electron
quasiparticles are in thermal equilibrium with the crystal in the
laboratory reference frame, so their distribution function is the
equilibrium Fermi function $n(E_k)$.  However, it is not
straightforward to apply the Fermi function, because the two
components of the eigenfunction (\ref{psi}), which have the same
energy in the reference frame of the moving CDW/SDW, have different
energies in the laboratory frame. Let us make a reasonable assumption
that a state (\ref{psi}) is populated according to its {\em average}
energy $\bar{E}_k^\pm$:
\begin{eqnarray}
  \bar{E}_k^\pm&=&|u_k^\pm|^2[\pm E_k+\hbar(k_{\rm F}+k)v]
\nonumber \\
  &&{}+|w_k^\pm|[\pm E_k+\hbar(-k_{\rm F}+k)v].
\label{E} 
\end{eqnarray}
The electric current $I$ carried by the electrons is equal to
\begin{eqnarray}
  &&I=2e\hbar\sum_{\pm}\int\frac{dk}{2\pi}\,n(\bar{E}_k^\pm) 
\label{I} \\
  && \times\left[|u_k^\pm|^2
  \left(\frac{k_F+k}{m}+\frac{v}{\hbar}\right) + |w_k^\pm|^2
  \left(\frac{-k_F+k}{m}+\frac{v}{\hbar}\right)\right]\!,
\nonumber
\end{eqnarray}
where the factor 2 comes from the spin. Substituting Eq.\ (\ref{E})
into Eq.\ (\ref{I}) and keeping the terms linear in $v$, we find two
contributions to $I$. The first contribution, $I_1$, is obtained by
replacing $\bar{E}_k^\pm$ with $\pm E_k$ in Eq.\ (\ref{I}), that is,
by omitting $v$ in Eq.\ (\ref{E}). This term represents the current
produced by all electrons moving with the velocity $v$:
\begin{equation}
I_1=2evk_F/\pi.
\label{I1}
\end{equation}
The second contribution $I_2$ comes from expansion of the Fermi
function in Eq.\ (\ref{I}) in $v$ and represents reduction of
the current due to thermally excited quasiparticles staying behind the
collective motion:
\begin{eqnarray}
  &&I_2=2emv\sum_{\pm}\int\frac{dk}{2\pi}\:
  \frac{\partial n(\pm E_k)}{\partial E_k}
\nonumber \\
  && \times\left(v_F(|u_k^\pm|^2-|w_k^\pm|^2)
  +\frac{\hbar k}{m}(|u_k^\pm|^2+|w_k^\pm|^2)\right)^2.
\label{I2}
\end{eqnarray}
The second term in the brackets in Eq.\ (\ref{I2}) is small compared
to the first term and may be neglected. Substituting Eqs.\ (\ref{uw1})
and (\ref{uw2}) into Eq.\ (\ref{I2}) and expressing the CDW/SDW
velocity in terms of the CDW/SDW phase derivative in time,
$v=-\dot{\tilde\Theta}/2k_F$, we find the temperature-dependent
expression for the Fr\"ohlich current:
\begin{equation}
  I=I_1+I_2=-f(T)\,\frac{e}{\pi}\frac{\partial\tilde\Theta}{\partial t},
\label{Iphi}
\end{equation}
\begin{equation}
  f(T)=1-\int d\xi_k
  \left(\frac{\xi_k}{E_k}\right)^{\!2}
  \left(-\frac{\partial n(E_k)}{\partial E_k}\right).
\label{ff(T)}
\end{equation}
Equation (\ref{ff(T)}) is the same as Eq.\ (\ref{f(T)}). Dividing the
current per one chain, $I$ (\ref{Iphi}), by the interchain distance
$b$, we get the density of current per unit length, $j_x$
(\ref{jxphi}).

\subsection{Diagrammatic derivation}
\label{sec:Matsubara}

In order to obtain the effective action for FISDW, we repeat the
derivation of Sec.\ \ref{sec:alternative} at a finite temperature.
Technically, this means that we need to calculate $\Pi^\mu$ in Eq.\ 
(\ref{eq:Pi_mu}) with $\Theta_0=0$ at the Matsubara frequencies
$i\omega_n=i(2n+1)\pi T/\hbar$, then make an analytic continuation to
the real frequencies and substitute the result into Eq.\ 
(\ref{eq:dS_Pi}):
\begin{eqnarray}
  &&\Pi^\mu(p,i\Omega_m)=-\frac{\tilde{\Delta}}{b}
  \int_{-\infty}^\infty \frac{dk}{2\pi}  
\label{jcorrelk} \\
  &&\times T\sum_n 
  {\rm Tr}[\tau_y{\cal G}(k,i\omega_n)\tau^\mu
  {\cal G}(k+p,i\omega_n+i\Omega_m)],
\nonumber
\end{eqnarray}
where the electron Green function is
\begin{equation}
  {\cal G}(k,i\omega_n)=
  \frac{\tau_0i\hbar\omega_n + \tau_z\hbar v_Fk + \tau_x\tilde\Delta}
  {(i\hbar\omega_n)^2-E_k^2}.
\label{GF2}
\end{equation}
Substituting Eq.\ (\ref{GF2}) into Eq.\ (\ref{jcorrelk}) and taking
the trace, we find
\begin{equation}
  \Pi^\mu(T,p,i\Omega_m)=
  -\varepsilon^{\mu\nu}\frac{ip_\nu}{\pi b}\,f(T,p,i\Omega_m), 
\label{eq:Pi}  
\end{equation}
where $p_\nu=(i\Omega_m/v_F,-p_x)$, and
\begin{eqnarray}
  &&f(T,p,i\Omega_m)=2\tilde\Delta^2\int_{-\infty}^\infty d\xi_k
\label{TrGG} \\
  &&\times T \sum_n \frac 1{[(i\hbar\omega_n)^2-E_k^2]
  [(i\hbar\omega_n+i\hbar\Omega_m)^2-E_{k+p}^2]}.
\nonumber 
\end{eqnarray}
The sum (\ref{TrGG}) is converted into an integral in the complex
plane of $\omega$ along a contour encircling the imaginary axis in the
counterclockwise direction with the function $-\hbar/T(2\pi
i)(e^{\hbar\omega/T}+1)$ multiplying the integrand in Eq.\ 
(\ref{TrGG}).  The integral is taken by deforming the contour of
integration into four contours encircling the four poles, $\pm
E_k/\hbar $ and $-i\Omega_m\pm E_{k+p}/\hbar $, in the clockwise
direction and evaluating the residues.  After that, we analytically
continue the external Matsubara frequency $i\Omega_m$ to the real
frequency: $i\Omega_m\to\Omega +i\delta$, where $\delta=\epsilon\,{\rm
  sign}(\Omega)$, and find
\begin{eqnarray}
  && f(T,p,\Omega)=\frac{1}{2}
   \int_{-\infty}^\infty d\xi_k\, \frac{\tilde{\Delta}^2}{E_kE_{k+p}} 
\label{ACcond} \\
  &&\times\left(
  \frac{n(E_{k+p})+n(E_k)-1}{\hbar\Omega +i\delta -E_k-E_{k+p} }
  -\frac{n(E_{k+p})+n(E_k)-1}{\hbar\Omega+i\delta +E_k+E_{k+p}}\right.
\nonumber \\
  &&+\left.\frac{n(E_{k+p})-n(E_k)}{\hbar\Omega +i\delta +E_k-E_{k+p}}
  -\frac {n(E_{k+p})-n(E_k)}{\hbar\Omega+i\delta -E_k+E_{k+p}}\right).
\nonumber
\end{eqnarray}
The second line in Eq.\ (\ref{ACcond}) contains the sum $E_k+E_{k+p}$
in the denominators and describes the interband electron transitions
involving the energy greater than $2\tilde{\Delta}$.  On the other
hand, the third line in Eq.\ (\ref{ACcond}) contains the difference
$E_k-E_{k+p}$ in the denominators and describes the intraband electron
transitions within the same energy band.

Substituting Eq.\ (\ref{eq:Pi}) into Eq.\ (\ref{eq:dS_Pi}) and taking
the variational integral over $\delta\tilde\Theta$, we find that the
effective action of FISDW at a finite temperature has the form
(\ref{eq:S_pa_2+1}) and (\ref{eq:L_eff_2+1}), Fourier-transformed from
$(t,x)$ to $(p,\Omega)$ and multiplied by the temperature-dependent
factor $f(T,p,\Omega)$ (\ref{ACcond}).  Since Lagrangian
(\ref{eq:L_eff_2+1}) represents an expansion in the powers of small
gradients, we would like to take the limit of $p\to0$ and $\Omega\to0$
in $f(T,p,\Omega)$.  However, at a finite temperature, the result
depends on the order of limits. In the dynamic limit, $\Omega\gg\hbar
v_Fp$, where we take the limit $p\to0$ before $\Omega\to0$, the
intraband cluster [the third line of Eq.\ (\ref{ACcond})] gives no
contribution, while the interband cluster (the second line) gives
\begin{eqnarray}
  f_d(T)&=&\lim_{\Omega\to0}\lim_{p\to 0}f(T,p,\Omega)
\nonumber \\
  &=&\int_{-\infty}^\infty d\xi_k
  \left(\frac{\tilde\Delta}{E_k}\right)^2
  \frac{1-2n(E_k)}{E_k}
\label{dynCond1'} \\
  &=&1-\int_{-\infty}^\infty d\xi_k
  \left(\frac{\xi _k}{E_k}\right)^2
  \left(-\frac{\partial n(E_k)}{\partial E_k}\right).
\label{dynCond1}
\end{eqnarray}
The integral of the first term in Eq.\ (\ref{dynCond1'}) gives 1 in
Eq.\ (\ref{dynCond1}), and the second term in Eq.\ (\ref{dynCond1'}),
integrated by parts, gives the second term in Eq.\ (\ref{dynCond1}).
The function $f_d(T)$ in the dynamical limit (\ref{dynCond1}) is the
same as the function $f(T)$ (\ref{ff(T)}) derived
semiphenomenologically in Sec.\ \ref{sec:Galileo}.  The dynamic limit
is appropriate for calculating electric conductivity, including the
Hall conductivity, when the electric field and FISDW are strictly
homogeneous in space ($p=0$), but may be time-dependent
($\Omega\neq0$).  Thus, the ac Hall conductivity at a finite
temperature is given by Eq.\ (\ref{eq:omega}) and Fig.\ 
\ref{fig:Hall(omega)} multiplied by $f(T)=f_d(T)$.

The function $f_s(T)$ in the static limit, $\Omega\ll\hbar v_Fp$, is
obtained from $f_d(T)$ (\ref{dynCond1}) by adding the intraband
contribution:
\begin{eqnarray}
  f_s(T) &=&\lim_{p\to0}\lim_{\Omega\to0}f(T,p,\Omega)
\nonumber \\
  &=&f_d(T) -\int_{-\infty}^\infty  d\xi_k
  \left(\frac{{\tilde{\Delta}}}{E_k}\right)^2 
  \left(-\frac{\partial n(E_k)}{\partial E_k}\right).
\label{statCond1}
\end{eqnarray}
Combining the second term in Eq.\ (\ref{dynCond1}) with the last term
in Eq.\ (\ref{statCond1}), we find
\begin{equation}
  f_s(T)=1-\int_{-\infty}^\infty d\xi_k 
  \left(-\frac{\partial n(E_k)}{\partial E_k}\right).
\label{statCond}
\end{equation}
The function $f_s(T)$ in the static limit is the same as the function
that determines the temperature reduction of the superfluid condensate
density in London superconductors, $\rho_s(T)/\rho$, \cite{Fetter71}.
This quantity controls the Meissner effect and, thus, determines the
temperature dependence of the magnetic field penetration depth in
superconductors. It also controls the charge-density response to a
static deformation of the CDW phase, $\partial\Theta/\partial x$
\cite{Lee79,Maki90}.  The static limit is appropriate in these cases,
because the CDW phase or the magnetic field in the Meissner effect are
stationary ($\Omega=0$), but vary in space ($p\neq0$).  Different, but
equivalent expressions for $f_d(T)$ and $f_s(T)$ were obtained in
Ref.\ \cite{Maki90} by integrating over the internal momentum of the
loop, $k$, in Eq.\ (\ref{ACcond}) first.

Comparing the definition (\ref{eq:Pi}) of the function $f$ with Eqs.\
(\ref{eq:Pi_mu}), (\ref{eq:dS_Q}), and (\ref{eq:QC}), we find that at
zero temperature $f=-C/2$, where $C$ is the Chern number.  At zero
temperature, the last terms in Eqs.\ (\ref{dynCond1}) and
(\ref{statCond}) vanish, so that $f_d(T=0)=f_s(T=0)=1$, which agrees
with the value $-2$ of the Chern number (\ref{eq:C}).  We may think of
$-2f(T)$ as a generalization of the Chern number to a finite
temperature, where it is not an integer topological invariant any
more, because of discrete summation, instead of integration, over the
frequency $\omega$.

The dependence of the function $f(T,p,\Omega)$ on the order of limits
indicates that the function is not analytic at small $p$ and $\Omega$.
Thus, at $T\neq0$, the effective Lagrangian of the system cannot be
written in a local form in the coordinate space as an expansion in
powers of gradients for an arbitrary relation between the time and
space gradients, so the momentum representation should be used.  For
the finite-temperature (2+1)D Chern-Simons theory this was emphasized
in Refs.\ \cite{Kao93,Aitchinson95}.  Another finite-temperature
effect is dissipation, which manifests itself as the imaginary part of
$f(T,p,\Omega)$ appearing at $\Omega\leq v_Fp$.  This Landau damping,
originating from the intraband electron transitions, is also known in
the theory of CDW/SDW \cite{Brazovskii93} and superconductivity
\cite{Abrahams66}.

\section{EXPERIMENTAL IMPLICATIONS} 
\label{sec:experiment}

In this paper, we predict two specific functions that can be measured
experimentally.  One is the frequency dependence (Fig.\ 
\ref{fig:Hall(omega)}) and another is the temperature dependence
(Fig.\ \ref{Fig:sxy(T)}) of the Hall conductivity.

The temperature dependence of the Hall {\it resistivity} in
(TMTSF)$_2$PF$_6$ was measured in experiments
\cite{Chaikin92e,Brooks96}.  However, to compare the experimental
results with our Eqs.\ (\ref{f(T)2Ne2/h})--(\ref{dynMaki}) for
$\sigma_{xy}(T)$, it is necessary to convert the Hall resistivity into
the Hall {\it conductivity}, which requires experimental knowledge of
all components of the resistivity tensor.  Only the temperature
dependences of $\rho_{xx}$ and $\rho_{xy}$, but not $\rho_{yy}$, were
measured in Refs.\ \cite{Chaikin92e,Brooks96}. Measuring the
temperature dependences of all three components of the resistivity
tensor and reconstructing $\sigma_{xy}(T)$ would play the same role
for QHE as measuring the temperature dependences of the magnetic-field
penetration depth for superconductors.

The frequency dependence of the Hall conductivity in regular
semiconductor QHE systems was measured using the technique of crossed
wave guides \cite{Kuchar,Galchenkov}. Unfortunately, no such
measurements were performed in the FISDW systems.  These measurements
would be very interesting, because they would reveal the competition
between the FISDW motion and QHE.  The required frequency should
exceed the FISDW pinning frequency $\omega_0$ and the damping rate
$1/\tau$.  To give a crude estimate of the required frequency range,
we quote the value of the pinning frequency $\omega_0\sim$ 3 GHz
$\sim$ 0.1 K $\sim$ 10 cm for a regular SDW (not FISDW) in
(TMTSF)$_2$PF$_6$ \cite{Quinlivan}.  One would expect a smaller value
for FISDW.

FISDW can be depinned not only by an ac electric field, but also by a
strong dc electric field.  The FISDW depinning and the influence of
steady FISDW sliding on the Hall effect were observed experimentally
in Refs.\ \cite{Osada87,Balicas93,Balicas98}.  Because the steady
sliding of a density wave is controlled by dissipation, it is
difficult to interpret these experiments quantitatively within a
microscopic theory.  According to our theory, in the dc case, the
nontrivial terms that couple the $x$ and $y$ directions along and
across the chains [the last terms in Eqs.\ (\ref{eq:jy}) and
(\ref{eq:fric}) proportional to $\ddot{\Theta}$ and $\dot{\cal E}_y$]
vanish.  Thus, the only effect of the FISDW sliding is an additional
Fr\"{o}hlich current along the chains, $\Delta j_x$, which is a
nonlinear function of ${\cal E}_x$.  In other words, the main effect
of the dc FISDW sliding would be a nonlinear increase in $\sigma_{xx}$
and, possibly, in $\sigma_{yy}$ via increasing the number of excited
nonequilibrium quasiparticles, but we would expect no major effect on
$\sigma_{xy}$.  Nevertheless, the dc FISDW sliding would affect the
experimentally measured Hall resistivity, because $\rho_{xy}$ depends
on all components of the conductivity tensor.

On a more subtle level, in the presence of a magnetic field, one could
phenomenologically add a term proportional to $E_y$ to Eq.\ 
(\ref{eq:fric}) and a term proportional to $\dot{\Theta}$ to Eq.\ 
(\ref{eq:jy}).  These terms would directly modify $\sigma_{xy}$ for
the dc sliding of FISDW.  Because these terms violate the
time-reversal symmetry of the equations, their nature must be
dissipative.  Thus, they cannot be derived within the Lagrangian
formalism, employed in this paper, and should be obtained from the
Boltzmann equation, where the time-reversal symmetry is already
broken.  The steady motion of the density-wave condensate itself does
not contribute to the Hall effect; however, this motion influences the
thermally excited normal carriers and, in this way, affects the Hall
voltage.  This picture is complimentary to our theory, which studies
only the condensate contribution.  Because the normal carriers need to
be thermally excited across the FISDW energy gap, we expect these
dissipative terms to be exponentially small and negligible at low
temperatures.

The influence of steady sliding of a regular CDW on the Hall
conductivity was studied theoretically in Ref.\ \cite{Artemenko84b}
along the lines explained in the preceding paragraph.  Since the bare
value of the Hall conductivity in a regular CDW/SDW system is
determined by the normal carriers only, the steady motion of the
density wave produces a considerable, of the order of unity, effect on
the Hall conductivity, which was observed experimentally
\cite{Artemenko84a}.  On the other hand, in the case of the FISDW,
where the big quantum contribution from the electrons below the gap
dominates the Hall conductivity, the contribution of the thermally
excited normal carriers to the Hall conductivity should be negligible
at low temperatures.

\section{CONCLUSIONS} 
\label{sec:conclusions}

In this paper, we have derived the effective Lagrangian
(\ref{eq:L_eff_2+1}), equivalently represented by Eq.\ (\ref{eq:CS'}),
for free FISDW.  The effective Lagrangian (\ref{eq:CS'}) consists of
the (2+1)D Chern-Simons term and the (1+1)D chiral-anomaly term, both
written for the effective field $A^i+a^i$, where $A^i$ is an external
electromagnetic field, and $a^i$ is the chiral field (\ref{eq:a_tx})
associated with the gradients of FISDW.  When FISDW is pinned, this
effective Lagrangian produces QHE.  On the other hand, in the ideal
case where FISDW is free, the counterflow of FISDW precisely cancels
the quantum Hall current, so the resultant Hall conductivity is zero.
The ac Hall conductivity $\sigma_{xy}(\omega)$ (\ref{eq:omega})
interpolates between these two limits at low and high frequencies, as
shown in Fig.\ \ref{fig:Hall(omega)}.

At a finite temperature, the effective Lagrangian (\ref{eq:L_eff_2+1})
or (\ref{eq:CS'}) should be multiplied by the function $f(T,p,\Omega)$
given by Eq.\ (\ref{ACcond}), which has the dynamic and static limits
$f_d(T)$ (\ref{dynCond1}) and $f_s(T)$ (\ref{statCond}).  The dynamic
limit determines the temperature dependence of the Hall conductivity,
which is given by Eqs.\ (\ref{f(T)2Ne2/h})--(\ref{dynMaki}) and shown
in Fig.\ \ref{Fig:sxy(T)}.  By analogy with the BCS theory of
superconductivity, this temperature dependence can be interpreted
within the two-fluid picture of QHE, where the Hall conductivity of
the condensate is quantized, but the condensate fraction of the total
electron density decreases with increasing temperature.

Experimentalists are urged to measure the frequency and temperature
dependences of $\sigma_{xy}$ in the FISDW state of the (TMTSF)$_2$X
materials.

\section*{ACKNOWLEDGMENTS}

V.M.Y.\ is grateful to S.~A.~Brazovskii and P.~B.~Wiegmann for useful
discussions.  This work was partially supported by the David and
Lucile Packard Foundation and by the NSF under Grant No.\ DMR-9417451.

\appendix

\section*{}
\label{sec:appendix}

In this appendix we calculate the Chern number (\ref{eq:C}):
\begin{equation}
  C=\int\frac{dk\,d\omega\,d\Theta_0}{4\pi^2}
  {\rm Tr}\left(
  \frac{\partial{\cal G}_0^{-1}}{\partial\Theta_0} {\cal G}_0
  \frac{\partial{\cal G}_0^{-1}}{\partial\omega} {\cal G}_0
  \frac{\partial{\cal G}_0^{-1}}{\partial k} {\cal G}_0\right).
\label{eq:C0}
\end{equation}
In Eq.\ (\ref{eq:C0}) we added the index 0 to the Green functions
${\cal G}_0$ in order to remind that they depend on the constant phase
$\Theta_0$ via the substitution
$\Delta\to\tilde\Delta\exp(-i\tau_z\Theta_0)$ in Eq.\ (\ref{eq:G_k}).
Now let us make a unitary transformation that eliminates the phase
$\Theta_0$ from the Green functions:
\begin{equation}
  U^+{\cal G}_0 U={\cal G},\quad U=e^{i\tau_z\Theta_0/2},
\label{eq:Theta_0}
\end{equation}
where ${\cal G}$ is given by Eq.\ (\ref{eq:G_k}) with
$\Delta\to\tilde\Delta$.  Substituting Eq.\ (\ref{eq:Theta_0}) into
Eq.\ (\ref{eq:C0}) and taking into account that $\partial{\cal
  G}^{-1}/\partial\omega=\hbar$, we find
\begin{equation}
  C=\hbar\int\frac{dk\,d\omega\,d\Theta_0}{4\pi^2}
  {\rm Tr}\left(U^+
  \frac{\partial(U{\cal G}^{-1}U^+)}{\partial\Theta_0} U{\cal G}{\cal G}
  \frac{\partial{\cal G}^{-1}}{\partial k}{\cal G}\right).
\label{eq:CU}
\end{equation}
Since ${\cal G}$ does not depend on $\Theta_0$, we need to
differentiate only the matrices $U$ and $U^+$ in Eq,\ (\ref{eq:CU}),
which gives the following two terms:
\begin{equation}
  C=i\frac{\hbar}{2}\int\frac{dk\,d\omega\,d\Theta_0}{4\pi^2}{\rm Tr}
  \left(\tau_z{\cal G}\frac{\partial{\cal G}^{-1}}{\partial k}{\cal G}
  -{\cal G}{\cal G}\frac{\partial{\cal G}^{-1}}{\partial k}\tau_z
  \right).
\label{eq:C2}
\end{equation}
The second term in Eq.\ (\ref{eq:C2}) is proportional to $P^{00}$
(\ref{eq:I1}) and vanishes according to Eq.\ (\ref{eq:I13=0}), whereas
the first term is proportional to $P^{11}$ (\ref{eq:Idk}).  Using
Eqs.\ (\ref{eq:dG}) and (\ref{eq:n+n-}), we find the value of $C$:
\begin{eqnarray}
  C&=&-i\frac{\hbar}{2}\int\frac{dk\,d\omega}{2\pi}{\rm Tr}
  \left(\tau_z\frac{\partial{\cal G}}{\partial k}\right)
\nonumber \\
  &=&\int dk\,\frac{\partial[n_+(k)-n_-(k)]}{\partial k}=-2.
\label{eq:Cn}
\end{eqnarray}
For spinless fermions, the number in Eq.\ (\ref{eq:Cn}) would be $-1$.



\begin{references}
\bibitem{indices} The subscript and superscript indices $(i,j,k)$
  should not be confused with imaginary number $i=\sqrt{-1}$, the
  current $j$, and the wave vector $k$.
\bibitem{Theta_y} More precisely, transformation
  (\ref{eq:psi_tilde_2+1}) should be done in two steps.  First we
  transform the fermions in the momentum representation:
  $\psi'(k_y)=\exp\{i(Nb\tau_z/2)[(e/\hbar c)A_y-k_y]\}\psi''(k_y)$.
  Then we go from the momentum to the coordinate representation in $y$
  and do the second transformation
  $\psi''(y)=\exp[i(Nb\tau_z/2)\Theta(y)]\tilde\psi(y)$.  In this
  way we can accommodate the dependence of $\Theta$ on $y$.
\bibitem{E_x^2} In our derivation, we neglected the terms proportional
  to ${\cal E}_x^2$, ${\cal E}_y^2$, and ${\cal H}_z^2$ in the
  effective Lagrangian (\ref{eq:L_eff_2+1}).  These terms describe
  polarizability and magnetic susceptibility, but are not essential
  for calculating conductivity.
\end{references}
\end{document}